\documentclass[12pt]{article}
\usepackage{latexsym}

\let\ssection=\section
\renewcommand{\section}{\setcounter{equation}{0}\ssection}

\newcommand\mathC{\mkern1mu\raise2.2pt\hbox{$\scriptscriptstyle|$}
        {\mkern-7mu\rm C}} % The complex numbers
\newcommand{\mathR}{{\rm I\! R}}         % The real numbers

\newcommand\bi{\begin{itemize}}
\newcommand\ei{\end{itemize}}
\newcommand\be{\begin{equation}}
\newcommand\ee{\end{equation}}

\newcommand{\dd}{\ensuremath{{\delta}}}

\begin{document}
\begin{titlepage}

\begin{center}
{\large\bf Against {\em Pointillisme} about Mechanics }
\end{center}

\vspace{0.2 truecm}

\begin{center}
        J.~Butterfield\footnote{email: jb56@cus.cam.ac.uk;
            jeremy.butterfield@all-souls.oxford.ac.uk}\\[10pt] All Souls College\\ %%@
Oxford OX1 4AL
\end{center}

\begin{center}
       5 December 2005: forthcoming in {\em British Journal for the Philosophy of %%@
Science} 
\end{center}

\vspace{0.2 truecm}

\begin{abstract}
This paper forms part of a wider campaign: to deny {\em pointillisme}. That is the %%@
doctrine that a physical theory's fundamental quantities are defined at points of %%@
space or of spacetime, and represent intrinsic properties of such points or %%@
point-sized objects located there; so that properties of spatial or spatiotemporal %%@
regions and their material contents are determined by the point-by-point facts.

More specifically, this paper argues against {\em pointillisme} about the concept %%@
of velocity in classical mechanics; especially against proposals by Tooley, %%@
Robinson and Lewis. A companion paper argues against {\em pointillisme} about %%@
(chrono)-geometry, as proposed by Bricker. 

To avoid technicalities, I conduct the argument almost entirely in the context of  %%@
``Newtonian'' ideas about space and time, and the classical mechanics of %%@
point-particles, i.e. extensionless particles moving in a void. But both the %%@
debate and my arguments carry over to relativistic  physics.

\end{abstract}

\end{titlepage}

\tableofcontents

\section{Introduction}\label{Intr}
This paper forms part of a wider campaign: to deny {\em pointillisme}. That is the %%@
doctrine that a physical theory's fundamental quantities are defined at points of %%@
space or of spacetime, and represent intrinsic properties of such points or %%@
point-sized objects located there; so that properties of spatial or spatiotemporal %%@
regions and their material contents are determined by the point-by-point %%@
facts.\footnote{I think David Lewis first used the art-movement's name as a vivid %%@
label for this sort of doctrine: a precise version of which he endorsed.}

I will first describe this wider campaign (Section \ref{widercpgn}). Then I will %%@
argue against {\em pointillisme} as regards the concept of velocity in classical %%@
mechanics (Sections \ref{sec;pismintr} and \ref{sssec;LewRob}). A companion paper %%@
(Butterfield 2006) argues against {\em pointillisme} about (chrono)-geometry. In %%@
both cases, the main debate is about whether properties of a point that are %%@
represented by vectors, tensors, connections etc. can be intrinsic to the point; %%@
typically, {\em pointillistes}  argue that they can be. In both papers, I  focus %%@
on contemporary {\em pointillistes} who try to 
reconcile {\em pointillisme} with the fact that vectorial etc. properties seem %%@
extrinsic to points and point-sized objects, by proposing some heterodox construal %%@
of the properties in question.
 
The concept of velocity in mechanics provides two  illustrations of the lure of %%@
{\em pointillisme}, and this tendency to reconcile it with vectorial properties by  %%@
reconstruing physical quantities.\\
\indent (a): Tooley and others argue that for the sake of securing that a %%@
particle's instantaneous velocity is intrinsic to it at a time, we should not %%@
construe velocity in the orthodox way as a limit of average velocities---but %%@
instead reconstrue it along lines they propose. Again, my view is that there is no %%@
need for such heterodoxy: instead, we can and should reject {\em pointillisme}: %%@
(Section \ref{sec;pismintr}).\\
\indent (b): Robinson and Lewis argue that for the sake of securing a perdurantist %%@
account of identity over time (persistence), we should postulate vectorial %%@
properties numerically equal to velocity, but free of velocity's presupposition of %%@
the notion of persistence.  I maintain that if we reject {\em pointillisme}, the %%@
perdurantist has no need of such novel properties: (Section \ref{sssec;LewRob}).\\
\indent In similar vein, Bricker (1993) proposes to reconcile {\em pointillisme} %%@
with modern geometry's need for vectorial and tensorial properties  by re-founding %%@
geometry in terms of non-standard analysis, which rehabilitates the traditional %%@
idea of infinitesimals  (Robinson 1996). In (Butterfield 2006), I reply that once %%@
the spell of {\em pointillisme} is broken, these heterodox foundations of geometry %%@
are unmotivated.

So in both papers I defend orthodoxy about the foundations of both geometry and %%@
mechanics. But I do not mean to be dogmatic. As to geometry, I of course agree %%@
that there are several heterodox mathematical theories of the continuum that are %%@
technically impressive and philosophically suggestive. Butterfield (2006) gives %%@
some references; but I do not discuss details, since these theories offer no %%@
support for my target, {\em pointillisme}. More precisely:  these theories  do not %%@
suggest that  fundamental quantities represent intrinsic properties of points or %%@
point-sized bits of matter; because either they  do not attribute such quantities %%@
to points, or they even deny that there are any points.\footnote{Broadly speaking, %%@
the second option seems more radical and worse for {\em pointillisme}; though in %%@
such theories, the structure of a set of points is often recovered by a %%@
construction, e.g. on a richly structured set of regions.} So the upshot is that %%@
although  I am open to suggestions about heterodox treatments of the continuum, %%@
these treatments do not support {\em pointillisme}. Accordingly, I find the %%@
philosophical doctrine of {\em pointillisme} an insufficient reason for rejecting %%@
the orthodox treatment. 

Similarly in this paper for mechanics: though with the difference that %%@
mathematicians have not developed a handful of heterodox theories of mechanics, or %%@
of velocity, as they have of the continuum---one of which Tooley and the other %%@
authors might hope to invoke, to provide their heterodox new foundations of %%@
mechanics.\footnote{As we shall see, the proposals by Tooley and others have %%@
slight mathematical aspects; but these aspects do not amount to any such theory.} %%@
So the upshot is the same as for geometry:  I will again find {\em pointillisme} %%@
an insufficient reason for rejecting orthodoxy---though I am open  to suggestions! %%@
In this spirit, I will end by offering a sort of peace-pipe to Robinson and Lewis. %%@
I will formulate a cousin of their proposal---a cousin that is not {\em %%@
pointilliste}---and compare it with their original (Section \ref{secIntro}).

I will conduct the discussion almost entirely in the context of  ``Newtonian'' %%@
ideas about  space and time, and the classical mechanics of point-particles, i.e. %%@
extensionless point-masses moving in a void (and so interacting by %%@
action-at-a-distance forces). This restriction keeps things simple: and at no %%@
cost,  since both the debate and my arguments carry over to relativistic physics. %%@
The restriction also has another merit. Broadly speaking, of the various physical %%@
theories, it is the classical mechanics of point-particles that {\em pointillisme} %%@
fits best: other theories have further anti-{\em pointilliste} features. So it is %%@
worth emphasising that even for classical point-particles, {\em pointillisme} %%@
fails.

\section{The wider campaign}\label{widercpgn}
As I mentioned, this paper is part of a wider campaign, which I now sketch. I %%@
begin with general remarks, especially about the intrinsic-extrinsic distinction %%@
among properties (Section \ref{ssec;connect}). Then I state my main claims; first %%@
in brief (Section \ref{ssec;apfp}), then in more detail (Section %%@
\ref{ssec;moredetail}).

\subsection{Connecting physics and metaphysics}\label{ssec;connect}
My wider campaign aims to connect what modern classical physics says about matter %%@
with two debates in modern analytic metaphysics. The first debate is about {\em %%@
pointillisme}; but understood as a metaphysical doctrine rather than a property of %%@
a physical theory. So, roughly speaking, it is the debate whether the world is %%@
fully described  by all the intrinsic properties of all the points and-or %%@
point-sized bits of matter. The second debate is whether an object persists over %%@
time by the selfsame object existing at different times (nowadays called  `{\em %%@
endurance}'), or by different temporal parts, or stages, existing at different %%@
times (called `{\em perdurance}').

Endeavouring to connect classical physics and metaphysics raises two large initial %%@
questions  of philosophical method. What role, if any, should the results of %%@
science have in metaphysics? And supposing metaphysics should in some way %%@
accommodate these results, the fact that we live (apparently!) in a  quantum %%@
universe prompts the question why we should take classical physics to have any %%@
bearing on metaphysics. I address these questions in my (2004: Section 2, 2006a: %%@
Section 2). Here I just summarize my answers.\\
\indent  I of course defend the relevance  of the results of science for %%@
metaphysics; at least for that branch of it, the philosophy of nature, which %%@
considers such notions as space, time, matter and causality. And this includes %%@
classical physics, for two reasons.\\
\indent First, much analytic philosophy of nature assumes, or examines, so-called %%@
`common-sense' aspects and versions of these notions: aspects and versions which %%@
reflect   classical physics, especially mechanics, at least as taught in %%@
high-school or elementary university courses. One obvious example is modern %%@
metaphysicians' frequent discussions of matter as point-particles, or as continua %%@
(i.e. bodies whose entire volume, even on the smallest scales, is filled with %%@
matter): of course, both notions arose in mechanics in the seventeenth and %%@
eighteenth centuries.\\
\indent Second, classical physical  theories, in particular mechanics, are much %%@
more philosophically suggestive, indeed subtle and problematic, than philosophers  %%@
generally realize. Again, point-particles and continua provide examples. The idea %%@
of  mass concentrated in a spatial point (indeed, different amounts at different %%@
points) is, to put it mildly, odd; as is action-at-a-distance interaction. And %%@
there are considerable conceptual tensions in the mechanics of continua; (Wilson %%@
(1998) is a philosopher's introduction). Unsurprisingly, these subtleties and %%@
problems were debated in the heyday of classical physics, from 1700 to 1900; and %%@
these debates had an enormous influence on philosophy through figures like Duhem, %%@
Hertz and Mach---to mention only figures around 1900 whose work directly %%@
influenced the analytic tradition. But after the quantum and relativity %%@
revolutions, foundational issues in classical mechanics were largely ignored, by %%@
physicists and mathematicians as well as by philosophers. Besides, the growth of %%@
academic philosophy after 1950 divided the discipline into  compartments, labelled %%@
`metaphysics', `philosophy of science' etc., with the inevitable result that there %%@
was less communication between, than within, compartments.\footnote{Thus I see my %%@
campaign as a foray into the borderlands between metaphysics and philosophy of %%@
physics: a territory that I like to think of as inviting exploration, since it %%@
promises to give new and illuminating perspectives on the theories and views of  %%@
the two communities lying to either side of it---rather than as a no-man's-land %%@
well-mined by two sides, ignorant and suspicious of each other!}

Setting aside issues of philosophical method, {\em pointillisme} and persistence %%@
are clearly large topics; and each is the larger for being treatable using the %%@
very diverse methods and perspectives of both disciplines, metaphysics and %%@
physics. So my campaign has to be selective in the ideas I discuss and in the %%@
authors  I cite. Fortunately, I can avoid several philosophical controversies, and %%@
almost all technicalities of physics.\footnote{I note that among the philosophical %%@
issues my campaign avoids are several about persistence, such as: (a) the gain and %%@
loss of parts (as in Theseus' ship); (b) the relation of ``constitution'' between %%@
matter and object (as in the clay and the statue); (c) vagueness, and whether %%@
there are vague objects. Agreed, there are of course  connections between my %%@
claims and arguments, and the various issues, both philosophical and physical, %%@
that I avoid: connections which it would be a good project to explore. But not in %%@
one paper, or even in one campaign!}

But it will clarify the purposes of this paper to give at the outset some details %%@
about how I avoid philosophical controversy about the intrinsic-extrinsic %%@
distinction among properties, and about how this distinction differs from three %%@
that are prominent in mathematics and physics.

\subsubsection{Avoiding controversy about the intrinsic-extrinsic %%@
distinction}\label{sssec;ied}
My campaign does not need to take sides in the ongoing controversy about how to %%@
analyse, indeed understand, the intrinsic-extrinsic distinction. (For an %%@
introduction, cf. Weatherson (2002, especially Section 3.1), and the symposium, %%@
e.g. Lewis (2001), that he cites.) Indeed, most of my discussion can make do with %%@
a much clearer distinction, between what Lewis (1983, p. 114) dubbed the `positive %%@
extrinsic' properties, and the rest. This goes as follows. \\
\indent Lewis was criticizing Kim's proposal, to analyze extrinsic properties as %%@
those that imply {\em  accompaniment}, where something is accompanied iff it %%@
coexists with some wholly distinct contingent object, and so to analyze intrinsic %%@
(i.e. not extrinsic) properties as those that are compatible with being %%@
unaccompanied, i.e. being the only contingent object in the universe (for short: %%@
being {\em lonely}). Lewis objected that loneliness is itself obviously extrinsic. %%@
He also argued that there was little hope of amending Kim's analysis. In %%@
particular, you might suggest that to be extrinsic, a property must either imply %%@
accompaniment or imply loneliness: so Lewis dubs these disjuncts `positive %%@
extrinsic' and `negative extrinsic' respectively. But Lewis points out that by %%@
disjoining and conjoining properties, we can find countless extrinsic properties %%@
that are neither positive extrinsic nor negative extrinsic; (though `almost any %%@
extrinsic property that a sensible person would ever mention is positive %%@
extrinsic' (1983, p. 115)).\\
\indent  This critique of Kim served as a springboard: both for Lewis' own %%@
preferred analysis, using a primitive notion of naturalness which did other %%@
important work in his metaphysics (Lewis 1983a); and for other, metaphysically %%@
less committed, analyses, developed  by Lewis and others (e.g. Langton and Lewis %%@
1998, Lewis 2001).\\
\indent But I will not need to pursue these details. As I said, most of my %%@
campaign can make do with the notion of positive extrinsicality, i.e. implying %%@
accompaniment, and its negation. That is, I can mostly take {\em pointillisme} to %%@
advocate properties that are intrinsic in the weak sense of not positively %%@
extrinsic. So this makes my campaign's claims, i.e. my denial of {\em %%@
pointillisme}, logically stronger; and so I hope more interesting. Anyway, my %%@
campaign (even in this paper) makes some novel proposals about positive %%@
extrinsicality: namely, I distinguish temporal and spatial (positive) %%@
extrinsicality, and propose degrees of (positive) extrinsicality.

\subsubsection{Distinction from three mathematical %%@
distinctions}\label{sssec;3maths}
Both the murky intrinsic-extrinsic distinction, and the clearer distinction %%@
between positive extrinsics and the rest, are different distinctions from three %%@
that are made within mathematics and physics, especially in those parts relevant %%@
to us: viz. pure and applied differential geometry. The first of these %%@
distinctions goes by the name `intrinsic'/`extrinsic'; the second is called %%@
`scalar'/`non-scalar', and the third is called `local'/`non-local'. They are as %%@
follows.

\indent (i): The use of `intrinsic' in differential geometry is a use which is %%@
common across all of mathematics: a feature is intrinsic to a mathematical object %%@
(structure) if it is determined (defined) by just the object as given, without %%@
appeal to anything extraneous---in particular a choice of a coordinate system, or %%@
of a basis of some vector space, or of an embedding of the object into another. %%@
For example, we thus say that the intrinsic geometry of a cylinder is flat; it is %%@
only as embedded in $\mathR^3$ that it is curved.

\indent (ii): Differential geometry classifies quantities according to how they %%@
transform between coordinate systems: the simplest case being scalars which have %%@
the same value in all coordinate systems. (Nevermind the details of how the other %%@
cases---vectors, tensors, connections, spinors etc.---transform.)

\indent (iii): Differential geometry uses `local' (as vs. `global') in various %%@
ways. But the central use is that a mathematical object (structure) is local if it %%@
is associated with a point by being determined (defined) by the mathematical  %%@
structures defined on {\em any} neighbourhood, no matter how small, of the point. %%@
In this way, the instantaneous velocity of a point-particle at a spacetime point, %%@
and all the higher derivatives of its velocity, are local since their existence %%@
and values are determined by the particle's trajectory in an arbitrarily small %%@
neighbourhood of the point. Similarly, an equation is called `local' if it %%@
involves only local quantities. In particular, an equation of motion is called %%@
`local in time' if it describes the evolution of the state of the system at time %%@
$t$ without appealing to any facts that are a finite (though maybe very small) %%@
time-interval to the past or future of $t$.

I will not spell out {\em seriatim} some examples showing that the two %%@
philosophical distinctions are different from the three mathematical ones. Given %%@
some lessons in differential geometry (not least learning to distinguish (i) to %%@
(iii) themselves!), providing such examples is straightforward work. Suffice it to %%@
make two comments; the second is relevant to this paper.

(1): It would be a good project to explore the detailed relations between these %%@
distinctions. In particular, the mathematical distinction (i) invites comparison %%@
with Vallentyne's (1997) proposal about the intrinsic-extrinsic distinction. %%@
Besides, there are yet other distinctions to explore and compare: for example, %%@
Earman (1987) catalogues some dozen senses of `locality'.

\indent (2): Instantaneous velocity, conceived in the orthodox way as a limit of %%@
average velocities, has implications about the object at other times, for example %%@
that it persists for some time.  (I will discuss this in more detail below, %%@
especially  Section \ref{ssec;vely?}.) So most philosophers say that instantaneous %%@
velocity is an extrinsic property.  I agree. But emphasising its extrinsicness %%@
tends to make one ignore the fact that it is mathematically local, i.e. determined %%@
by the object's trajectory in an arbitrarily small time-interval (cf. (iii) %%@
above). It is this locality  that prompts me to speak of instantaneous velocity %%@
(and other local quantities) as `hardly extrinsic'. And in pure and applied %%@
differential geometry, it would be hard to over-estimate the importance of---and %%@
practitioners' preference for!---such local quantities and local equations %%@
involving them. Similarly, the fact that we often find that differential equations %%@
of very low order determine the temporal course of quantities of interest, is very %%@
important---and very fortunate.\footnote{Of course, we sometimes need equations of %%@
higher order than we at first think and hope. For an important case of this in %%@
population biology, cf. Colyvan and Ginzburg (2003).}

\subsection{Classical mechanics is not {\em pointilliste}, and can be  %%@
perdurantist}\label{ssec;apfp}
\subsubsection{Two versions of {\em pointillisme}}\label{sssec;2versions}
To state my campaign's main claims, it is convenient to first distinguish a weaker %%@
and a stronger version of {\em pointillisme}, understood as a metaphysical %%@
dosctrine. They differ, in effect, by taking `point' in {\em pointillisme} to %%@
mean, respectively, spatial,  or spacetime,  point.

 Taking `point' to mean `spatial point', I shall take {\em pointillisme} to be, %%@
roughly, the doctrine that the instantaneous state of the world is fully described  %%@
by all the intrinsic properties, at that time, of all spatial points and-or %%@
point-sized bits of matter.\\
\indent As I said in Section \ref{ssec;connect}, my campaign can mostly take %%@
`intrinsic' to mean `lacking implications about some wholly distinct contingent %%@
object'; in other words, to mean the negation of Lewis' `positive extrinsic' (i.e. %%@
his `implying accompaniment'). But for this version of {\em pointillisme}, I will %%@
take `intrinsic' to mean `{\em spatially} intrinsic'. That is,  attributing such a %%@
property to an object carries no implications about spatially distant objects; but %%@
it {\em can} carry implications about objects at other times. (Such objects might %%@
be other temporal parts of the given object.) So I shall call this version, `{\em %%@
pointillisme} as regards space'.

On the other hand: taking `point' to mean `spacetime point', I shall take {\em %%@
pointillisme} to be, roughly, the doctrine that the history of the world is fully %%@
described  by all the intrinsic properties of all the spacetime  points and-or all %%@
the intrinsic properties at all the various times of point-sized bits of matter %%@
(either point-particles, or in a continuum). And here I take `intrinsic' to mean %%@
just the negation of Lewis' `positive extrinsic'. That is, it means `both %%@
spatially and temporally intrinsic': attributing such a property carries no %%@
implications about objects at other places, or at other times. I shall call this %%@
stronger version, `{\em pointillisme} as regards spacetime'.

So to sum up: {\em pointillisme} as regards space vetoes spatial extrinsicality; %%@
but {\em pointillisme} as regards spacetime also vetoes temporal extrinsicality.

On either reading of {\em pointillisme}, it is of course a delicate matter to %%@
relate such metaphysical doctrines, or the endurance-perdurance debate,  to the %%@
content of specific physical theories. Even apart from Section %%@
\ref{ssec;connect}'s questions  of philosophical method, one naturally asks for %%@
example, how  philosophers' idea of  intrinsic property relates to the idea of a %%@
physical quantity. For the most part, I shall state my verdicts about  such %%@
questions case by case. But one main tactic for relating the metaphysics to the %%@
physics will be to formulate {\em pointillisme} as a doctrine relativized to (i.e. %%@
as a property of) a given physical theory (from Section \ref{ssec;moredetail} %%@
onwards). Anyway, I can already state my main claims, in terms of these two %%@
versions of {\em pointillisme}. More precisely, I will state them as denials of %%@
two claims that are, I think,  common in contemporary metaphysics of nature.

\subsubsection{Two common claims}\label{sssec;2claims}
Though I have not made a survey of analytic metaphysicians, I think many of them %%@
hold two theses, which I will dub (FPo) (for `For {\em Pointillisme}') and (APe) %%@
(for `Against perdurantism'); as follows.

\indent (FPo): Classical physics---or more specifically, classical %%@
mechanics---supports {\em pointillisme}: at least as regards  space, though  %%@
perhaps not as regards spacetime. There are two points here:---\\
\indent \indent (a): Classical physics is free of various kinds of  ``holism'', %%@
and thereby anti-{\em pointillisme}, that are suggested by quantum theory. Or at  %%@
least: classical  mechanics is free. (With the weaker claim, one could allow, and %%@
so set aside, some apparently anti-{\em pointilliste} features of advanced %%@
classical physics, e.g. anholonomies in electromagnetism and the %%@
non-localizability of gravitational energy in general relativity: features rich in %%@
philosophical suggestions (Batterman 2003, Belot 1998, Hoefer 2000)---but not for %%@
this paper!)\\
\indent \indent (b): The concession, `perhaps  not as regards spacetime', arises %%@
from the endurance-perdurance debate. For it seems  that {\em pointillisme} as %%@
regards spacetime must construe persistence as perdurance;  (while {\em %%@
pointillisme} as regards space could construe it as endurance). And a well-known %%@
argument, often called `the rotating discs argument', suggests that perdurance  %%@
clashes with facts about the rotation of a continuum (i.e. a continuous body) in %%@
classical mechanics. So the argument suggests that classical mechanics must be %%@
understood as ``endurantist''. Besides, whether or not one endorses the argument,  %%@
in classical mechanics the persistence  of objects surely {\em can} be understood %%@
as endurance---which conflicts with {\em pointillisme} as regards spacetime.\\
\indent  (The considerations under (a) and (b) are usually taken as applying %%@
equally well to non-relativistic and  relativistic classical mechanics: an %%@
assumption I largely endorse.)

I also think that many metaphysicians would go further and hold that:\\
\indent (APe): Classical mechanics does indeed exclude {\em pointillisme} as %%@
regards spacetime: their reason being that this {\em pointillisme} requires %%@
perdurance and that they endorse the rotating discs argument. So they hold that in %%@
classical mechanics the persistence  of objects  {\em must} be understood as %%@
endurance, and that this forbids {\em pointillisme} as regards spacetime. 

\subsubsection{My contrary claims}\label{sssec;my2claims}
I can now state the main position of my wider campaign. Namely, I {\em deny} both %%@
claims, (FPo) and (APe), of Section \ref{sssec;2claims}. I  argue for two contrary %%@
claims, (APo) (for `Against {\em Pointillisme}) and (FPe) (for `For %%@
perdurantism'), as follows.

\indent (APo): Classical mechanics does {\em not} support {\em pointillisme}.\\
\indent By this I do not mean just that:\\
\indent\indent (a) it excludes {\em pointillisme} as regards spacetime.\\
 Nor do I just mean:\\
\indent\indent (b) it allows one to construe the persistence  of objects as %%@
endurance.\\
 (But I agree with both (a) and (b).)  Rather, I also claim: classical mechanics  %%@
excludes {\em pointillisme} as regards space. That is: it needs to attribute %%@
spatially extrinsic properties to spatial points, and-or point-sized bits of %%@
matter. (But this will not be analogous to the kinds of ``holism'' suggested by %%@
quantum theory.)
 
\indent (FPe): Though (as agreed in (APo)) classical mechanics  excludes {\em %%@
pointillisme} as regards spacetime (indeed, also: as regards space): classical %%@
mechanics is {\em compatible} with perdurance. That is: despite the rotating discs %%@
argument,  one {\em can} be a ``perdurantist'' about the persistence of objects in %%@
classical mechanics. The reason is that once we reject {\em pointillisme}, %%@
perdurance does not need persistence to supervene on temporally intrinsic facts. %%@
In fact, perdurantism can be defended by swallowing just a small dose of temporal %%@
extrinsicality. 
  
So to sum up my wider campaign, I claim that:---\\
\indent (APo): Classical mechanics denies {\em pointillisme}, as regards space as %%@
well as spacetime. For it needs to use spatially extrinsic properties of spatial %%@
points and-or point-sized bits of matter, more than is commonly believed. \\
\indent (FPe): Classical mechanics permits perdurantism. It does not require %%@
temporally extrinsic properties (of matter, or objects), in the sense of requiring %%@
persistence to be endurance: as is commonly believed. A mild dose of temporal %%@
extrinsicality can reconcile classical mechanics with perdurance.

To put the point in the philosophy of mind's terminology of `wide' and `narrow' %%@
states, meaning (roughly) extrinsic and intrinsic states, respectively: I maintain %%@
that classical mechanics:\\
\indent (APo): needs to use states that are spatially wide, more than is commonly %%@
believed; and \\
\indent (FPe): does not require a specific strong form of temporal width, viz. %%@
endurance. With a small dose of temporal extrinsicality, it can make do with %%@
temporally quite narrow states---and can construe persistence as perdurance.

\subsection{In more detail ...}\label{ssec;moredetail} 
So much by way of an opening statement.  I will now spell out my main claims in a %%@
bit more detail: (APo) in Section \ref{sssec;fourviol} and (FPe) in Section %%@
\ref{sssec;forperdm}.  

\subsubsection{Four violations of {\em pointillisme}}\label{sssec;fourviol}
I will begin by stating {\em pointillisme}  as a trio of claims that apply to any %%@
physical theory; and making two comments. Then I list four ways in which %%@
(chrono)-geometry and classical  mechanics violate {\em pointillisme}: three will %%@
form the main topics of this paper and its companion.

 The trio of claims is as follows:\\
\indent (a): the fundamental quantities of the physical theory in question are to %%@
be defined at points of space or of spacetime;\\
\indent (b): these quantities represent intrinsic properties of such points; \\
\indent (c): models of the theory---i.e. in physicists' jargon, solutions of its %%@
equations, and in metaphysicians' jargon, possible worlds according to the %%@
theory---are fully defined by a specification of the quantities' values at all %%@
such points.\\
\indent So, putting (a)-(c) together: the idea of {\em pointillisme} is that the %%@
theory's  models (or solutions or worlds) are something like conjunctions or %%@
mereological fusions of ``ultralocal facts'', i.e. facts at points.

Two comments. First: the disjunction in (a), `at points of space or of spacetime', %%@
corresponds to   Section \ref{ssec;apfp}'s distinction between {\em pointillisme} %%@
as regards space, and as regards spacetime. Nevermind that it does not imply the %%@
convention I adopted in Section \ref{ssec;apfp}, that {\em pointillisme} as %%@
regards spacetime is a {\em stronger} doctrine since it vetoes temporally %%@
extrinsic properties, {\em as well as} spatially extrinsic ones. The context will %%@
always make it clear whether I mean space or spacetime (or both); and whether I %%@
mean spatially or temporally extrinsic (or both).\\
\indent Second: Though I have not made a systematic survey, there is no doubt that %%@
{\em pointillisme}, especially its claims (a) and (b), is prominent in recent %%@
analytic metaphysics of nature, especially of neo-Humean stripe. The prime example %%@
is the metaphysical system of David Lewis, which is so impressive in its scope and  %%@
detail.  One of his main metaphysical theses, which he calls `Humean %%@
supervenience', is a version of {\em pointillisme}. I will return to this in %%@
Section \ref{sssec;LewRob}. 

When we apply (a)-(c) to classical mechanics, there are, I believe, four main ways %%@
in which {\em pointillisme} fails: or more kindly expressed, four concessions %%@
which {\em pointillisme} needs to make. The first three violations (concessions) %%@
occur in the classical mechanics both of point-particles and of continua; the %%@
fourth is specific to continua. The first two violations are discussed in the %%@
companion paper (2006); the third is the topic of this paper.

(1): The first is obvious and minor. Whether matter is conceived as %%@
point-particles or as continua, classical  mechanics uses a binary relation of %%@
occupation, `... occupies ...', between bits of matter and spatial or spacetime %%@
points (or, for extended parts of a continuum: spatial or spacetime regions). And %%@
this binary relation presumably brings with it extrinsic properties of its relata: %%@
it seems an extrinsic property of a point-particle (or a continuum, i.e. a %%@
continuous body) that it occupy a certain spatial or spacetime point or region; %%@
and conversely.

\indent (2): Classical mechanics (like other physical theories) postulates %%@
structure for space and-or spacetime (geometry or chrono-geometry); and this %%@
involves a complex network of geometric  relations between, and so extrinsic %%@
properties of, points. This concession is of course more striking as regards space %%@
than time: three-dimensional Euclidean geometry involves more structure than does %%@
the real line. This is the main topic of (2006).

\indent (3): Mechanics needs of course to refer to the instantaneous  velocity or %%@
momentum of a body; and this is temporally extrinsic to the instant  in question, %%@
since for example it implies the body's existence at other times. (But it is also %%@
local in the sense of (iii), Section \ref{sssec;3maths}.)  So this second %%@
violation imposes temporal, rather than spatial, extrinsicality; i.e.  %%@
implications about other times, rather than other places.

This is the main topic of this paper. But I should stress that this third %%@
violation is {\em mitigated} for point-particles. For a {\em pointilliste} {\em %%@
can} maintain that the persistence of point-particles supervenes on facts that, %%@
apart from the other violations (i.e. about `occupies' and (chrono)-geometry), are %%@
{\em pointillistically} acceptable: viz. temporally intrinsic facts about which %%@
spacetime  points are occupied by matter. In figurative terms: the void between %%@
distinct point-particles allows one to construe their persistence in terms of %%@
tracing the curves in spacetime connecting points that are occupied by matter. I %%@
develop this theme in my (2005). On the other hand: for a continuous body, the %%@
persistence of spatial parts (whether extensionless or extended) does {\em not} %%@
supervene on such temporally intrinsic facts. This is the core idea of  the %%@
rotating discs argument, mentioned in Section \ref{sssec;2claims}.\\
\indent To sum up: the rotating discs argument means that {\em pointillisme} fits %%@
better with point-particles than with continua. To put the issue in terms of %%@
Section \ref{ssec;apfp}'s two forms of {\em pointillisme}: the strong form of {\em %%@
pointillisme}, {\em pointillisme} as regards spacetime, fails for the classical %%@
mechanics of continua, even apart from the other concessions mentioned. 

(4): Finally, there is a fourth way that the classical mechanics of continua %%@
violates {\em pointillisme}: i.e., a fourth concession that {\em pointillisme} %%@
needs to make. Unlike the rotating discs argument, this violation seems never to %%@
have been noticed in recent analytic metaphysics; though the relevant physics goes %%@
back to Euler. Namely, the classical mechanics of continua violates (the weaker %%@
doctrine of) {\em pointillisme} as regards space, because it must be formulated in %%@
terms of spatially extended regions and their properties and relations. But in %%@
this paper, I set this fourth violation aside entirely; my (2006a) gives details.
  
So to sum up these four violations, I claim (APo): classical mechanics violates %%@
{\em pointillisme}. This is so even for the weaker doctrine, {\em pointillisme}  %%@
as regards space. And it is especially so, for the classical mechanics of continua %%@
rather than point-particles.

\subsubsection{For perdurantism}\label{sssec;forperdm}
I turn to Section \ref{sssec;my2claims}'s second claim, (FPe):  that once {\em %%@
pointillisme} is rejected, perdurantism does not need persistence to supervene on %%@
temporally intrinsic facts, and can be defended for classical mechanics provided %%@
it swallows a small  dose of temporal extrinsicality. 

Now I can identify this small dose. It is the extrinsicality of Section %%@
\ref{sssec;fourviol}'s third violation of {\em pointillisme}; in particular,  the %%@
presupposition of persistence by the notion of a body's instantaneous velocity. %%@
Thanks to the rotating discs argument, `body' here means especially `point-sized %%@
bit of matter in a continuum'. For as we noted in Section \ref{sssec;fourviol}, %%@
for point-particles we can construe persistence as perdurance without having to %%@
take this dose. 

Elsewhere (2004, 2004a) I argue that for a ``naturalist'' perdurantist, this dose %%@
is small enough to swallow. For this paper (especially Section \ref{sssec;LewRob}) %%@
I only need to state the argument's two main ideas:\\
\indent (i): If the perdurantist rejects {\em pointillisme}, she can reject %%@
instantaneous temporal parts, i.e. believe only in temporal parts with some %%@
non-zero duration.\\
\indent (ii): She can thereby avoid the rotating discs argument. For the argument %%@
urges that facts temporally intrinsic to instants cannot distinguish two obviously %%@
different states of motion for a continuous body. But the anti-{\em pointilliste} %%@
perdurantist has access to non-instantaneous facts, and so can ``thread the %%@
worldlines together''.

But I should  also stress that I do not claim to {\em refute} endurantism, even %%@
for so limited and sharply-defined a class of objects  as the point-particles and %%@
continua of classical mechanics. The metaphysical debate about persistence  is  %%@
too entangled with other debates in the philosophy of time, and in ontology and %%@
semantics, for me to claim that. I claim only that in classical mechanics at %%@
least, perdurantism is tenable. In fact, I think that in classical mechanics, the %%@
cases for endurantism and perdurantism are about equally strong: the honours are %%@
about even. That is an ecumenical conclusion---but one worth stressing since for %%@
continua, perdurantism has got such a bad press, thanks to the rotating discs %%@
argument.

\section{Velocity as intrinsic?}\label{sec;pismintr}

\subsection{Can properties represented by vectors be intrinsic to a %%@
point?}\label{ssec;pismprospect}
Classical mechanics represents  the properties that encode the structure of space %%@
or spacetime, and the properties  of matter such as velocity, momentum etc., using %%@
mathematical entities such as vectors, tensors, connections etc. (Of course, so do %%@
all physical theories.) So the question arises: can properties that are so %%@
represented  be intrinsic to a point? This question is central to {\em %%@
pointillisme}, and to our other topic, persistence: and will be at the centre of %%@
this paper.

But my discussion will be simplified by two drastic restrictions. First, I will %%@
consider only properties represented by vectors, which I will for short call {\em %%@
vectorial properties}: not those represented by other mathematical entities such %%@
as tensors and connections. Though drastic, this restriction is natural, in %%@
that:\\
\indent (i) vectors are about the simplest of the various mathematical entities %%@
that classical mechanics (like other theories) uses to represent properties and %%@
relations---so they are the first case to consider;\\  
\indent (ii) the restriction is common in the literature: of the authors I discuss %%@
in this paper and its companion, all consider only vectorial properties, except %%@
for Bricker (1993) who also considers tensors.

Second, I will concentrate on instantaneous velocity. For the {\em pointilliste} %%@
authors I will criticize (mainly Tooley, Robinson and Lewis) do so; though both %%@
they and I will also briefly comment on momentum, force and acceleration.

As announced in Section \ref{Intr}, the discussion will illustrate how strongly %%@
some contemporary metaphysicians are attracted by {\em pointillisme}. They %%@
reconcile the apparent extrinsicality of a vectorial property, specifically %%@
velocity, with {\em pointillisme} by proposing to reconstrue the property. Thus in %%@
Section \ref{sssec;Tooley}, Tooley and others will reject the orthodox idea of  %%@
instantaneous velocity as a limit of average velocities, and reconstrue it in %%@
order to make it an intrinsic property. And in Section \ref{sssec;LewRob}, %%@
Robinson and Lewis will make a similar proposal. My own view will of course be %%@
that there is no need for such heterodoxy: instead, we can and should reject {\em %%@
pointillisme}.

So the plan of battle for the rest of this paper is as follows. I will undertake %%@
four projects: the first two in this Section, and then two in Section %%@
\ref{sssec;LewRob}:---\\
\indent (1): I will  endorse the view that instantaneous velocity is extrinsic, in %%@
particular temporally extrinsic. But I will also emphasise that because it is %%@
local ((iii) of Section \ref{sssec;3maths}), it is hardly extrinsic; (Section %%@
\ref{ssec;vely?}).\\
\indent (2): I will criticize the heterodox view of Tooley and others that we %%@
should reconstrue velocity so as to make it intrinsic; (Section %%@
\ref{sssec;Tooley}).\\
\indent (3): I will criticize the view of Lewis and Robinson that a moving object %%@
has a vectorial property numerically equal to velocity, but free of velocity's %%@
presupposition of the notion of persistence; (Sections \ref{sssec:LRproposal} and %%@
\ref{sssec;LRassessed}).\\
\indent (4): I end by offering a peace-pipe to Robinson and Lewis. I formulate a %%@
cousin of their proposal---a cousin that is not {\em pointilliste}---and end by %%@
comparing it with their original; (Section \ref{secIntro}).\\
\indent So while the second and third projects are critical, the first and fourth %%@
are more positive. In particular, Section \ref{secIntro} reflects Section %%@
\ref{Intr}'s admission that heterodox treatments of the continuum and of physical %%@
quantities are of course worth developing.

\subsection{Orthodox  velocity is extrinsic but local}\label{ssec;vely?}
\subsubsection{A question and a debate}\label{qndebate}
To lay out the ground, let us begin by asking: Is a particle's velocity intrinsic %%@
to it at the time in question? The first thing to say is of course  that 
this question, and similar ones e.g. whether a particle's velocity counts as part %%@
of its instantaneous state, have a long history. Although `intrinsic' is a %%@
philosophical term of art (especially nowadays, cf. Section \ref{sssec;ied}), and %%@
although instantaneous velocity was first rigorously defined only with the advent %%@
of the calculus,  the vaguer notion of ``velocity at a time'' was  involved in all %%@
debate about the nature of motion from Zeno's time onwards.

I will not go in to details details about this long history. Here it must suffice %%@
to say that its ``highlights'' include: Zeno, Ockham's `at-at' theory of motion, %%@
medieval impetus theory, the apparent resolution of Zeno's paradoxes provided by %%@
the calculus (at least as rigorized by Cauchy and Weierstrass, if not by its %%@
seventeenth century inventors), the philosophical discussion of that resolution by %%@
analytic philosophers like Russell, and the recent development, in mathematics, of %%@
heterodox theories of the continuum---for example, the  two  modern vindications %%@
of the idea of infinitesimals, non-standard analysis and smooth infinitesimal %%@
analysis. (Sources include: for the history, Mancosu (1996, Chapters 4f.), Leibniz %%@
(2001), Arthur (2006); for the recent work on infinitesimals, Robinson (1996), %%@
Bell (1998).)

So I set aside both the history and contemporary infinitesimals; and I postpone  %%@
heterodox philosophical treatments of velocity to the next Subsection. Here I will %%@
give the details of the orthodox answer to our question. Namely, as  announced in %%@
Section \ref{sssec;fourviol}: instantaneous velocity is extrinsic, in particular %%@
temporally extrinsic, but local. Though this answer is, I submit, straightforward, %%@
and the underlying mathematics is elementary, it is worth pausing over; for it has %%@
been the topic of some recent debate (between Albert, Arntzenius and Smith). 

So let us consider the orthodox notion of instantaneous velocity for a %%@
point-particle. This is the limit of the particle's average velocity as the %%@
time-interval around the point in question  tends to zero. To be precise, we will %%@
require the two one-sided limits and the two-sided limit to all exist and be %%@
equal: we say, in an obvious notation,
\begin{eqnarray}
{\rm If} \;\; {\rm lim}_{\varepsilon \rightarrow 0+} \frac{{\bf q}(t + %%@
\varepsilon) - {\bf q}(t)}{\varepsilon} = 
{\rm lim}_{\varepsilon \rightarrow 0+} \frac{{\bf q}(t - \varepsilon) - {\bf %%@
q}(t)}{\varepsilon} =  \nonumber \\
{\rm lim}_{\varepsilon \rightarrow 0+, \dd \rightarrow 0+} \frac{{\bf q}(t + %%@
\varepsilon) - {\bf q}(t - \dd)}{\varepsilon + \dd}, \;\;\; 
{\rm then} \;\; {\bf v}(t) := {\rm this \;\; common \;\; limit}.
\label{eq;defineinstsvely}
\end{eqnarray}
If we now ask our question---is velocity, defined by eq. \ref{eq;defineinstsvely}, %%@
intrinsic to the particle at $t$?---intuition pulls in two directions. 

On the one hand, one wants to say Yes, because an attribution of velocity (even of %%@
a specific value ${\bf v}$) implies no categorical information about the %%@
particle's velocity, or location, at other times. Having instantaneous velocity %%@
${\bf v}$ at a point ${\bf q}$ at time $t$ is compatible with {\em any} values for %%@
the particle's  instantaneous velocity and its location, at any other  instant %%@
$t'$ in time, as near as you please to $t$. Indeed in classical mechanics, the %%@
compatibility is not just logical, but also nomic, since classical mechanics gives %%@
no upper bound to either the speed or the acceleration of a particle.\\
\indent To put the same point in other words: there is no time-interval around $t$ %%@
(in particular, no minimal time-interval) for which the course of values of the %%@
particle's location and-or velocity, or some proposition about the possibilities %%@
for these courses of values,  are equivalent to the particle's instantaneous %%@
velocity at $t$. So the instantaneous velocity is surely not a property of the %%@
particle during a time-interval around $t$---which suggests that it is intrinsic %%@
to the particle  at $t$.  
 
But on the other hand, one wants to say No, for two reasons. First, the particle's %%@
velocity is relative to a frame of reference. This surely makes it  a relation %%@
between  the particle and an object stationary in (or perhaps in some other way %%@
representing) the frame. (Or at least, the velocity is an extrinsic property that %%@
the particle has because of such a relation.) Though this reason is important, it %%@
tends to be ignored in discussions about whether velocity is intrinsic; so it will %%@
be clearest to postpone it to Section \ref{sssec;commonview}.B, where I briefly %%@
discuss the two authors who mention it.

The second reason for answering No {\em is} recognized in the literature. It is %%@
that the particle's instantaneous velocity ${\bf v}$ at $t$ codes a lot of %%@
information about what its velocity and location is at nearby times---but not %%@
``categorical information''. The information is conditional or hypothetical %%@
information about average velocities (and consequently, locations). The %%@
information is given precisely by eq. \ref{eq;defineinstsvely}. And it is given %%@
roughly, by saying that for nearby times the collection of average velocities must %%@
be so ``well-behaved'' as to have a single limit, ${\bf v}$, as the times get %%@
closer to $t$; or in spacetime terms, by saying that the nearby history of %%@
locations (the local segment of the worldline) must be smooth enough to have at %%@
$t$ a tangent vector (a 4-velocity determined by ${\bf v}$).  

These diverse intuitions are clearly in evidence in the recent debate between %%@
Albert, Arntzenius  and Smith. I will not arbitrate this debate in detail: that %%@
would require extended quotation and textual exegesis. But  a short summary will %%@
suffice to  bring out the diverse intuitions; and I submit, to make clear that the %%@
right answer to our question is `No: velocity is extrinsic though local'.

The debate runs as follows:---\\
\indent (i): Albert (2000, pp. 9-10, 17-18) and Arntzenius (2000, Section 3, %%@
especially pp. 192-195) do not address exclusively our question. Their attention %%@
is predominantly on the similar question, whether a  particle's instantaneous %%@
velocity as defined by eq. \ref{eq;defineinstsvely} should count as part of its %%@
instantaneous state. They argue that it should not, since they require that an %%@
object's instantaneous state should not imply, in virtue of logic and definition %%@
alone, any constraints on its instantaneous states at other times. And velocity %%@
clearly does imply such constraints: roughly, that for some time-interval around %%@
$t$, maybe very short, the particle's average velocities are suitably %%@
``well-behaved''; cf. the `No' answer above. \\
\indent (ii): Smith (2003, pp. 269-280; especially pp. 274-277) replies that %%@
instantaneous velocity {\em should} count as part of the instantaneous state at %%@
$t$, essentially because for any {\em other} time the state (in particular the %%@
location and velocity) could be anything; cf. the `Yes' answer %%@
above.\footnote{Smith (pp. 264-268) also corrects the common view that Russell in %%@
his (1903) took the calculus to vindicate the idea of instantaneous velocity as %%@
intrinsic to the  object at the time. In fact, Russell's version of the ``at-at'' %%@
theory of motion denies that there are instantaneous states of motion (and more %%@
generally: of change). Tooley (1988: Sections 1, 2.2) makes the same point.} \\
\indent (iii): In his brief reply, Arntzenius (2003) emphasises that his and %%@
Albert's main idea is the requirement reported in (i), that an instantaneous state %%@
should  not imply, by just logic and definitions, any constraints on instantaneous %%@
states at other times. And he argues that velocity's implications about other %%@
times surely make it extrinsic. \\
\indent (iv): In a yet briefer rejoinder (2003a), Smith: (a) is sceptical about %%@
metaphysicians' intrinsic-extrinsic distinction; and (b) rejects Albert's and %%@
Arntzenius' requirement, since `physicists have not chosen to adhere to that %%@
requirement, and in the absence of a good reason ... we ought to stick with %%@
physics here' (2003a, p. 283).

\subsubsection{The verdict}\label{verdict}
Surveying this debate, I think the verdict is clear. Ultimately, it is of course %%@
just a verbal matter whether to impose Albert's and Arntzenius' requirement as %%@
part of the meaning of `instantaneous state'; though {\em ceteris paribus}, one is %%@
well-advised to follow the usage of the discipline concerned---and so, in this %%@
case, to join the physicists in not imposing it.\\
\indent But similarly, `intrinsic' and `extrinsic' are established philosophical %%@
terms. And there is no doubt that an attribution of instantaneous velocity (either %%@
the determinable or a determinate value) has implications for other times (not %%@
least that the object exists then), and so is extrinsic, indeed positive %%@
temporally extrinsic; (cf. Section \ref{sssec;ied} and the `No' answer above). In %%@
short: once the meaning of `extrinsic' is settled, it is uncontentious that %%@
velocity is extrinsic.\\
\indent On the other hand, the ``grain of truth'' in the `Yes' answer is that %%@
velocity is local in the sense of  (iii) Section \ref{sssec;3maths}: whether the %%@
particle has a velocity at $t$, and if so what it is, is determined by its %%@
positions at times in an arbitrarily short time interval around $t$. Again this is %%@
uncontentious.\footnote{But terminology varies. Bricker (1993, p. 289) calls such %%@
a property `neighbourhood-dependent'; similarly, Arntzenius (2000, p. 193) %%@
suggests calling them `neighbourhood properties'.}\\
\indent One might add to this verdict that the physicists' usage, that velocity is %%@
part of the instantaneous state, fits dynamics, as well as kinematics (i.e. as %%@
well as velocity's being local). For in a deterministic theory like mechanics, the %%@
laws of motion are to determine all later states from the present state (`initial %%@
data'); and since these laws are second-order in time, the initial data must %%@
include the velocity, as well as the position.\footnote{This point is not specific %%@
to Newton's laws, with force proportional to acceleration $d^2 q /dt^2$. Suitably %%@
generalized (to substitute momentum for velocity), it applies to both Lagrangian %%@
and Hamiltonian formulations of classical mechanics; and to relativistic %%@
generalizations.}    

Though uncontentious, this verdict is important for the debate about whether %%@
persistence is endurance or perdurance; and in particular, for my pro-perdurantist %%@
claim (FPe) (Section \ref{sssec;forperdm}). Thus one way in which velocity is %%@
extrinsic is its implication that the object exists at other times; and this has %%@
prompted a consensus that the perdurantist's reply to the rotating discs argument %%@
cannot appeal to different velocities (or angular velocities). But extrinsicality, %%@
though usually discussed as an all-or-nothing affair, comes in degrees (Lewis %%@
1983, p. 111):  a property is more extrinsic, the more that its ascription implies %%@
about the world beyond the property's instance. Since velocity is local, it is %%@
hardly extrinsic; and this means that a perdurantist who swallows this small dose %%@
of temporal extrinsicality can invoke velocity in her reply to the rotating discs %%@
argument. More precisely, an anti-{\em pointilliste} perdurantist who vetoes %%@
instantaneous temporal parts can do so. (Butterfield (2004, Sections 4.2.2 and %%@
7.4; 2004a, Sections 2.2.2, 4.5) gives details.)

Finally, two technical remarks which will be needed in Section \ref{sssec;Tooley}. %%@
(1): In eq. \ref{eq;defineinstsvely} and the ensuing discussion, we have %%@
implicitly  assumed that the particle exists throughout a time interval around %%@
$t$, not least because textbooks of analysis define continuity and %%@
differentiability at a point $t$ only for functions defined on a neighbourhood of %%@
$t$. But philosophers, concerned with logical as well as nomic possibilities, %%@
sometimes consider particles that come in and out of existence (e.g. Tooley and %%@
others discussed in Section \ref{sssec;Tooley}). So it is worth noticing that the %%@
usual definitions of continuity and differentiability at a point $t \in \mathR$ %%@
can be carried over to any function  defined on a subset of $\mathR$ that has $t$ %%@
as a limit point from both above and below. For example, one can talk about the %%@
continuity and differentiability at zero of a function defined on zero together %%@
with the reciprocals of integers, i.e. defined on $\{ 1/n \; : \; n \; {\rm{an}} %%@
\; {\rm{integer}} \; \} \cup \{ 0 \}$.\\
\indent (2): The orthodox account of velocity can also be liberalized, as regards %%@
one-sided limits. There is no reason to insist that a ``rate of motion'' deserves %%@
the name `velocity' only if both one-sided limits exist and are equal to the %%@
two-sided limit (cf. eq. \ref{eq;defineinstsvely}). Thus it is common practice to %%@
call the limit from above (eq. \ref{eq;defineinstsvely}'s first term) $v_+(t)$; %%@
and to call the limit from below (eq. \ref{eq;defineinstsvely}'s second term) %%@
$v_-(t)$; and to talk of one-sided velocities in a case where $v_+(t) \neq %%@
v_-(t)$. Combining this idea with (1) above, we see that a one-sided limit at a %%@
point $t$ only requires the function's domain to have $t$ as a limit point from %%@
that side.   

So much for the orthodox notion of velocity. From now on, I will consider, and %%@
rebut, heterodox views:---\\
\indent (1): in Section \ref{sssec;Tooley}, a view of velocity as intrinsic, which %%@
has been advocated without regard to the debate about persistence; and\\
\indent (2): from Section \ref{sssec;LewRob}, a view that accepts orthodox %%@
velocity but proposes that to understand persistence we need to postulate another %%@
vectorial quantity, always equal in value to orthodox velocity, but free of its %%@
presupposition of persistence.

\subsection{Against intrinsic velocity}\label{sssec;Tooley}
\subsubsection{A common view---and a common problem}\label{sssec;commonview}
\paragraph{3.3.1.A The view}\label{331Aview}
Several authors have sketched a heterodox view of instantaneous  velocity as %%@
intrinsic: Tooley (1988, p. 236f.), Bigelow and Pargetter (1989, especially pp. %%@
290-294; 1990, pp. 62-82)\footnote{I will only cite the former, since the latter %%@
is almost identical.} and Arntzenius (2000: pp. 189, 196-201). As we shall see, a %%@
dozen or so philosophers have commented on this heterodox view, and often %%@
sympathetically. But so far as I know, these authors' arguments for their view %%@
have not received detailed scrutiny---or rebuttal. So that will be my purpose %%@
until Section \ref{sssec;LewRob}. 

These authors' proposals seem to be mutually independent: the three later authors %%@
do not cite the previous work. But they share a common view, as follows:\\
\indent (i): Velocity should be an intrinsic property of an object at a time  that %%@
(together with the position, and the regime of impressed forces) causes, and so %%@
explains, the object's position at later times. \\
\indent (ii): Causation should be understood in a broadly neoHumean way, as a %%@
contingent relation between `distinct existences'. This means that the orthodox %%@
notion of velocity, being  a ``logical construction'' out of the object's %%@
positions at other times (cf. Section \ref{ssec;vely?}) cannot do the job required %%@
by (i). 

But the proposals differ in detail. For example, Tooley presents his proposal  by %%@
applying Lewis' (1970) tactic for functional definition of theoretical terms to a %%@
version of ``Newton's laws of motion''. Or to use another jargon: he ``Ramsifies'' %%@
some accepted laws of motion. I presume that by thus ``piggy-backing'' on the %%@
accepted laws, Tooley's approach can readily secure that its intrinsic velocity is %%@
a vector; (though Tooley does not go into this).  On the other hand, Bigelow and %%@
Pargetter develop the view without using details about the laws of motion: %%@
instead, they appeal to the metaphysics of universals and the logic of relations %%@
to argue that their intrinsic velocity is vectorial.\\
\indent Another difference concerns how to treat vectorial quantities other than %%@
velocity. Tooley proposes to treat force in a similar way to velocity but is %%@
content with an orthodox account of acceleration as ``just'' $d^2 {\bf q}/dt^2$ %%@
(1988, p. 249). But Bigelow and Pargetter (1989, pp. 294-295) propose heterodoxy %%@
for both force and acceleration; though not for higher derivatives of position, %%@
which, they assert, play no explanatory role.\footnote{For all three, the shared %%@
view (i) and (ii) exemplifies a characteristically Australian realism, and %%@
endorsement of inference to the best explanation. Thus Tooley elsewhere proposes a %%@
functional definition of causation (1987, Chapters 1.2, 8); and even proposes that %%@
a spacetime point causes later ones, and that in the context of relativity, this %%@
is a linear non-branching relation, providing a version of relativity theory with %%@
an absolute simultaneity (1997, pp. 338-344, 354-355: for discussion, cf. Dainton %%@
2001, pp. 278-281). Bigelow and Pargetter elsewhere defend a realist view of %%@
forces as a species of causation (Bigelow, Ellis and Pargetter, 1988).} \\
\indent There are also differences as regards connections to other topics. Thus %%@
Bigelow, Pargetter and Arntzenius, but not Tooley, suggest the view is a %%@
descendant of some medieval views (about impetus and flux).  And Arntzenius (2000, %%@
pp. 198-201) connects the view to time-reversal, and so to Albert's heterodox %%@
allegation (2000, pp. 14-15, 20-21) that electromagnetism is not time-reversal %%@
invariant.\footnote{Earman (2002), Arntzenius (2004) and Malament (2004) are %%@
replies to Albert.} 

In this Subsection, I will focus on Tooley, and so speak of `Tooleyan velocities'. %%@
My  reasons for this focus are that:---\\
\indent (i): So far as I know, his arguments for the view are the most developed; %%@
but as I mentioned, they seem not to have received detailed scrutiny. \\
\indent (ii): This focus enables me to avoid Bigelow and Pargetter's contentious %%@
metaphysical territory of universals.\footnote{Besides, I will reply {\em en %%@
passant} to their three principal arguments for distinguishing intrinsic velocity %%@
from orthodox velocity, since they are similar to arguments of Tooley's which I  %%@
discuss. Another discussion of Tooleyan velocities which emphasises causation more %%@
than I will, but whose verdict, like mine, is broadly negative, is Le Poidevin %%@
(2006).}

\paragraph{3.3.1.B The problem}\label{331Bproblem}
The view that velocity is intrinsic faces an obvious problem, which I mentioned in %%@
Section \ref{qndebate}.  Namely, velocity being intrinsic conflicts with velocity %%@
being relative to a frame of reference.  For the latter surely makes velocity  a %%@
relation between objects, viz. the given one and an object stationary in (or in %%@
some other way representing) the frame; or perhaps, a corresponding extrinsic %%@
property: in any case, not intrinsic.

Among advocates of the view, only Tooley, so far as I know, addresses this %%@
problem; and among commentators, only Zimmerman (1998, pp. 276-277). In Section %%@
\ref{sssec;Tooleymore},  I shall criticize Tooley's response to this problem; %%@
which is in any case brief. Here I want just to emphasise two points which force %%@
the problem on {\em any} advocate of intrinsic velocity.

\indent (1): The problem is not specific to special relativity; (as Tooley's %%@
response and Zimmerman's discussion both suggest). Classical mechanics no less %%@
than relativity can be formulated without postulating absolute rest, so that %%@
velocity is indeed relative to a frame of reference. (For the idea, think of the %%@
galilean transformations. For a rigorous formulation along these lines, one uses a %%@
neoNewtonian conception of spacetime: for philosophical expositions cf. e.g. Sklar %%@
(1974: Chapter III.D.3, pp. 202-206), Earman (1989, Chapter 2.4, p. 33).) 

\indent (2): The problem takes us back to the distinction between the %%@
philosophical and the mathematical notions of intrinsic, where the latter means in %%@
particular, coordinate-independent (cf. (i) of Section \ref{sssec;3maths}). To be %%@
precise, this distinction clarifies the problem---and shows that  any advocate of %%@
intrinsic velocity faces it. Thus it is not enough for the advocate to point out %%@
that rigorous formulations of mechanics, whether classical or relativistic, %%@
associate a mathematically intrinsic notion of 4-velocity to an object such as a %%@
point-particle, viz. the tangent vector to the object's worldline. For that fact %%@
does not imply that the 4-velocity is philosophically intrinsic, and thereby fit %%@
(by the advocate's lights) to do the jobs of causing and explaining later %%@
positions.  And even if the 4-velocity is philosophically intrinsic, that would %%@
not imply that a 3-velocity, i.e. an ordinary spatial velocity of the sort all the %%@
advocates discuss, is philosophically intrinsic and so ``fit for work''. For to %%@
define the 3-velocity from the 4-velocity, one has  to choose a frame of %%@
reference---in either a neoNewtonian or a relativistic theory: and so face the %%@
problem.\footnote{These comments apply equally to classical and relativistic %%@
mechanics. For the present topic, they only differ in the metrical properties %%@
attributed to a 4-velocity: in a neoNewtonian theory it has only a temporal %%@
length, while in a relativistic theory, it has a spatiotemporal length. %%@
Incidentally, Zimmerman's comment  (1998, pp. pp. 276-277) amounts to: (i) my (2), %%@
but applied only to relativity theory; (ii) the suggestion that 4-acceleration is %%@
a `much better candidate for an intrinsic state of motion' (p. 267). Le Poidevin %%@
(2006: Section 6) also makes the suggestion (ii). I agree with (ii), not least %%@
because in both theories, 4-acceleration has a coordinate-independent length. But %%@
it would take us too far afield to assess whether Tooley and his ilk could or %%@
should adapt the proposal and arguments in Section \ref{sssec;Tooleyalone} to %%@
acceleration instead of velocity: their project, not mine! Similarly, as regards %%@
adapting them to 4-velocity, rather than 3-velocity.}

So this problem is substantial. Nevertheless, metaphysicians seem to still %%@
consider the view  a live option; (e.g. Zimmerman (1998, pp. 275-278) and Sider %%@
(2001, pp. 35, 39, 228). And I will also set the problem aside, apart from briefly %%@
reporting Tooley's response (in Section \ref{sssec;Tooleymore}). After all, the %%@
view is so far only a sketch: none of these authors makes their  heterodox account %%@
of velocity part of a mathematically elaborated theory of motion (and-or %%@
causation).\footnote{Tooley (1988, pp. 231-2) briefly considers whether his %%@
account could be formalized using non-standard analysis' infinitesimals; but %%@
concludes that it cannot be.} But my setting this problem aside is not just an act %%@
of charity. There will be plenty else to comment on---and criticize! 

\subsubsection{Tooley's proposal; and his arguments}\label{sssec;Tooleyalone}
\paragraph{3.3.2.A Tooley's proposal}\label{332A;Tooleypropose}
As I mentioned, Tooley proposes that velocity is an intrinsic property that is %%@
functionally defined by the laws of motion, in the manner of Lewis (1970: %%@
especially Section IV). So, roughly speaking: Tooley says that the velocity of %%@
object $o$ at time $t$ is that unique intrinsic property of $o$ at $t$ that is %%@
thus-and-thus related to other concepts, as spelled out in the usual formulas of %%@
kinematics and dynamics. More precisely, Tooley gives a kinematic formula and a %%@
dynamical one, which he labels $T_1$ and $T_2$, Writing $q(t), v(t)$ for the %%@
position  and velocity at $t$ of the object $o$ with mass $m$, and $F(t)$ for the %%@
force impressed on $o$ at $t$, these formulas are:\footnote{Cf. Tooley (1988, pp. %%@
238-239). For brevity, I have suppressed universal quantifiers and a variable for %%@
the object $o$; and I have simplified $T_2$ so as to assume a constant mass; %%@
Tooley's $T_2$ tries to accommodate relativity's velocity-dependence of mass by %%@
writing $m(t)$. Both Tooley and I simplify to one spatial dimension: the %%@
generalization to three spatial dimensions, ${\bf q}(t), {\bf v}(t)$ etc. is %%@
trivial.}
\begin{eqnarray}
T_1 \; : \;\;\; q(t_2) = q(t_1) + \int^{t_2}_{t_1} v(t) \; dt ; \\
T_2 \; : \;\;\; v(t_2) = v(t_1) + \int^{t_2}_{t_1} F(t)/m \; dt .
\label{eqT1T2}
\end{eqnarray}
So velocity  is to be implicitly defined in Ramsey-Lewis style, as the unique %%@
intrinsic property with the functional role enjoyed by the term $v$ in $T_1 \& %%@
T_2$.

As Tooley interprets this proposal, it differs both mathematically and %%@
philosophically from the orthodox account of eq. \ref{eq;defineinstsvely}. Let us %%@
first address the mathematical difference; which is minor.

Eq. \ref{eq;defineinstsvely} requires $q$ to be differentiable. But $T_1 \& T_2$ %%@
(indeed $T_1$) implies only that $q$ is continuous (since the integral %%@
$\int^{t_2}_{t_1} v \; dt$ is a continuous function of its limits)\footnote{Tooley %%@
makes a slip here (p. 238), saying that $T_1 \& T_2$ do not even imply that $q$ is %%@
continuous: no matter.}; not that it is differentiable, since $v$ in $T_1$ is {\em %%@
not} defined as $\frac{dq}{dt}$. Thus Tooley considers the case of a particle %%@
initially at rest at the origin, with $v(t) = 0$ for all $t < 0$ and $v(t) = 1$ %%@
for all $t > 0$, so that $q(t) = 0$ for all $t < 0$ and $q(t) = t$ for all $t > %%@
0$. Because of the corner at $t = 0$, $q$ is not differentiable at $t = 0$; and %%@
however we might choose to define $v(0)$, $v$ will be discontinuous at 0. Yet %%@
$T_1$ holds good: in particular, $v$ in integrable. \\
\indent There are good mathematical questions hereabouts. For example: how %%@
``well-behaved'' must $v$ be in order to be integrable (on either the Riemann or %%@
the Lebesque definition)? And as a consequence: how well-behaved must the integral %%@
i.e. $q$ be? These questions are addressed in integration theory. But Tooley does %%@
not pursue them;\footnote{But Tooley does deploy the above example in one of his %%@
philosophical arguments for his proposal; cf. (2) in Section %%@
\ref{sssec;Tooleyalone}.B below.} and nor will I. For us it is enough to note that %%@
Tooley's $T_1 \& T_2$ is a mathematically mild generalization of eq. %%@
\ref{eq;defineinstsvely}'s orthodox account: in short, the difference is that %%@
whereas orthodoxy takes position as primitive and velocity as its derivative, $T_1 %%@
\& T_2$ takes velocity  as primitive and position as its integral.   

\paragraph{3.3.2.B Tooley's arguments}\label{332B;Tooleyargts}
Turning to philosophy, Tooley gives six arguments for his proposal, which together %%@
bring out how he interprets it; (his Sections 4.1-4.6, with further discussion and %%@
replies in Sections 5 and 6). He admits that the arguments vary in strength, and %%@
that they need a variety of deniable premises. These premises are typical of %%@
contemporary analytic metaphysics of nature. They concern such topics as:\\
\indent (a) whether there can be ``action at a temporal  distance'': i.e. roughly, %%@
a cause at $t_1$ of an effect at $t_2$, without causally relevant states of %%@
affairs at all the times between $t_1$ and $t_2$; or\\
\indent (b) whether motion can be discontinuous.\\
And as one might expect,  the arguments also depend on less explicit, but again %%@
deniable, premises (``intuitions''), articulating a broadly neoHumean view of %%@
causation; (cf. (ii) in Section \ref{sssec;commonview}.A).  

I will not try to state all Tooley's arguments; but will concentrate on the %%@
arguments and premises that seem most important. (This will cover three arguments %%@
given by Bigelow and Pargetter; cf. footnote 14.) This means I will focus on the %%@
arguments of his Sections 4.2, 4.4, 4.5 and 4.6. But it will be clearer to discuss %%@
his 4.5 before his 4.4; for 4.4 and 4.6 share a common premise, that motion could %%@
be discontinuous. It will also be clearest to reply to the arguments {\em %%@
seriatim}, so as to avoid having to refer back to arguments.\\
\indent But I can already state the general tenor of my reply. I will criticize %%@
Tooley's (and Bigelow and Pargetter's) arguments as either:\\
\indent (i): not justifying a crucial premise, or ``intuition'' about a %%@
thought-experiment; and-or\\
\indent (ii) under-estimating the resources of the orthodox account. \\
I would add that (i) and (ii) both arise from the arguments and %%@
thought-experiments being by and large too far removed from the details of %%@
mechanics. But of course a philosopher of physics {\em would} say that to a  %%@
metaphysician!

(1): {\em Tooley's Section 4.2}:--- \\
The argument of Tooley's Section 4.2 assumes a denial of ``action at a temporal  %%@
distance'' ((a) above), which he calls the `principle of causal continuity'. We do %%@
not need the exact formulation of the principle, but only this consequence of it:
\begin{quote}
If there is a causally sufficient condition of some state of affairs, however %%@
complex the condition and however gappy it may be [i.e. spread across disconnected %%@
intervals or instants of time---JNB], there must also be some instantaneous state %%@
of affairs which is also a causally sufficient condition of the state of affairs %%@
in question. (p. 242) 
\end{quote}
Tooley accepts that the principle of causal continuity is not a necessary truth, %%@
but holds that it is `reasonable' to believe it true of `our world' (1988, p. %%@
242). Presumably he would say the same about this consequence. In any case, he  %%@
then says
\begin{quote}
In either a Newtonian or a relativistic world ... the state of the world at an %%@
instant cannot be a causally sufficient condition of later states unless velocity %%@
(or, alternatively, something to which velocity is definitionally related, such as %%@
momentum) characterizes the instantaneous states of objects. If therefore the %%@
principle of causal continuity is accepted, the Russellian [i.e. orthodox] %%@
analysis of velocity must be rejected. (p. 242) 
\end{quote}
In the first sentence here, Tooley is of course referring to the fact noted in %%@
Section \ref{ssec;vely?}'s verdict on the Albert-Arntzenius-Smith debate: that %%@
since in Newtonian or relativistic mechanics the laws of motion are second-order %%@
in time, the initial data of a solution must include the velocity or momentum, so %%@
that it is natural to call velocity part of the instantaneous state. So far, so %%@
uncontentious: at least in so far as we go along with Tooley in regarding the %%@
determination of the later state by the present one (the initial data) as a case %%@
of causal sufficiency.\\
\indent But  the second sentence {\em is} contentious. Here, Tooley assumes that %%@
an orthodox velocity cannot count as part of the instantaneous state---at least if %%@
`instantaneous state' is understood as  causally sufficient for a later state.  %%@
Thus his view is like that of Arntzenius and Albert (cf. Section %%@
\ref{ssec;vely?}), though more explicit and more detailed in its commitment to a %%@
neoHumean view of causation. So my reply is: why should we accept this assumption?  %%@
So far, I see no reason: especially in the light of  Section \ref{ssec;vely?}'s %%@
verdict that orthodox velocity, though  extrinsic, is local and part of the %%@
instantaneous state.\footnote{The same reply works against Bigelow and Pargetter's %%@
similar, but more free-wheeling, argument. They go so far as to say that the %%@
extrinsicality of  orthodox velocity amounts to action at a temporal distance! %%@
Thus they allege that the orthodox description of an impact, e.g. a meteor %%@
striking Mars, requires the meteor's past positions to exert a force now; so they %%@
remark incredulously  that `this requires the meteor to have a kind of %%@
`memory'--what it does to Mars depends not only on its current properties  but %%@
also on where it has been' (1989, p. 296). I submit that Section %%@
\ref{ssec;vely?}'s discussion and verdict scotches this argument.}

(2): {\em Tooley's Section 4.5}:--- \\
The argument of Tooley's Section 4.5 uses Section \ref{sssec;Tooleyalone}.A's %%@
example of a particle which is at rest and then moves with velocity $v(t) = 1$ at %%@
all $t > 0$. Tooley argues that though orthodoxy dictates that $v(0)$ is %%@
undefined, the intuitive verdict about the case is that $v(0) = 0$. For since (as %%@
he argues elsewhere: 1987, pp. 207-212) cause and effect cannot be simultaneous, %%@
the motion's cause, viz. an instantaneous impulsive force acting at $t = 0$, can %%@
only have an effect (viz. $ v = 1$) later. Tooley then also assumes that his %%@
functional definition of velocity using $T_1 \& T_2$ will imply this result, i.e. %%@
that if the particle's movement is due to an impulsive force acting %%@
instantaneously, then $v(0) = 0$ (p. 246, paragraph 4). Finally, he says that in a %%@
world in which impulsive forces act ``at a temporal distance'', e.g. with a time %%@
delay of one second, and in which an appropriate\footnote{Namely, the mass of the %%@
particle times 1 unit of velocity.} instantaneous impulse is impressed on the %%@
particle at time $t = -1$ s., the velocity at $t = 0$ {\em would} be 1: $v(0) = %%@
1$. And again he assumes that his functional definition will imply this result (p. %%@
246-247).

\indent I reply that while Tooley's judgments about the intuitive values of %%@
velocity, in the light of various postulated causal stories, may well be %%@
defensible in an elaborated theory of causation and motion, they are debatable (i) %%@
in general and (ii) in the context of his $T_1 \& T_2$.\\
\indent (i): Philosophers with other views of causation might well disagree. And %%@
not only philosophers who  are sceptical about causal talk: philosophers who %%@
``believe in'' causation, but see little connection between causation and %%@
classical mechanics, in particular velocity, might well disagree.\\
\indent (ii):  In particular, these judgments do not just follow from $T_1 \& %%@
T_2$, or from a functional definition obtained from them. For the transition from %%@
orthodoxy, eq. \ref{eq;defineinstsvely}, to treating velocity as primitive and %%@
position as its integral, i.e. $T_1 \& T_2$, cannot settle disputed matters of %%@
causation---causal relations are too misty a subject to be settled by such a %%@
mathematically mild generalization.\footnote{
Bigelow and Pargetter's argument for Tooleyan velocity differing from orthodox %%@
velocity is different. They present a case where both are defined but differ, for %%@
an instant, in value (1989, pp. 292-293). Suppose two perfectly rigid spheres, B %%@
and C, are at rest and touching; then B is struck along the line between their %%@
centres by a third such sphere, A, at velocity $v$. Supposing the spheres are of %%@
equal mass, `theory tells us that A will stop, B will not budge, and C will move %%@
off with velocity $v$' (1989, p. 293); (think of a ``Newton's cradle''). Bigelow %%@
and Pargetter assert that at the instant of impact, B has a Tooleyan velocity $v$: %%@
`the velocity of A is transferred from A, through B, to C' (ibid.).\\
\indent I reply that this is a case where everyday or philosophers' intuitions are %%@
too far removed from the details of mechanics: in two ways.\\
\indent (i): That `A will stop, B will not budge, and C will move off' is only an %%@
approximate description, based on the idealization of perfect rigidity. This is a %%@
very strong idealization: by assuming forces, and finite impulses, are transmitted %%@
instantaneously through bodies, it forbids any account of processes within bodies. %%@
And this can be philosophically misleading; for example, in the rotating disc %%@
argument (Butterfield 2004: Section 5.5.2; 2004a: Section 3.2), and in the %%@
metaphysics of causation (Wilson 2004).\\
\indent (ii): Even if we assume perfect rigidity, it does not follow that B has a %%@
velocity, even a heterodox one for just an instant. It only follows that a finite %%@
impulse is transmitted through B.}

The arguments of Tooley's Sections 4.4. and 4.6 both assume that a particle's %%@
spatial trajectory could be discontinuous (not just non-differentiable) so that it %%@
``jumps about'' in space.

(3): {\em Tooley's Section 4.4}:--- \\
In Section 4.4, Tooley adds to discontinuous motion the idea of a world in which a %%@
particle's present position gives only probabilities for its later positions, and %%@
then asks us to consider a particle with just happens to have a spatial trajectory %%@
that is throughout some time interval a differentiable function of time. In short, %%@
we are to consider what Tooley dubs `accidentally orderly movement in a %%@
probabilistic world' (p. 243).\\
\indent  Tooley now urges the intuition that such a particle would {\em not} have %%@
a velocity at any time in the interval in question, because `the velocity of an %%@
object at a time should be {\em causally} relevant to its positions at later %%@
times' (p. 244).\\
\indent  Tooley also points out that the same example threatens any view that %%@
takes velocity to be determined by (supervenient upon) the history of positions; %%@
e.g. a liberalization of eq. \ref{eq;defineinstsvely} which required only that the %%@
one-sided limit, from earlier times, of average velocities, should exist. For the %%@
accidentally orderly history of position in Tooley's imagined probabilistic world %%@
could  match exactly a particle's history in a deterministic (say, classical %%@
mechanical) world. And the latter particle, says Tooley, {\em does} have a %%@
velocity---it is part of the instantaneous state and causally relevant to later %%@
positions. So if we accept Tooley's intuition that the accidentally orderly %%@
particle lacks a velocity, then the worlds match as to the particles' positions  %%@
but differ as to their velocity: and not just as to what the value of velocity is, %%@
but as to whether there is a value.

I make two replies, analogous to those for Tooley's Section 4.5 ((2) above). %%@
First,  intuitions clash.  I have no hesitation in judging, {\em contra} Tooley, %%@
that the particle moving accidentally in an orderly way has a velocity---even if I %%@
try to forget my previous inclination to the orthodox view! Nor am I alone; cf. %%@
Smith (2003, p. 279  fn 14). So far, this is a stalemate. But second, as in %%@
(2)(ii) above: Tooley's $T_1 \& T_2$, and a functional definition obtained from %%@
them, do not imply that the accidentally ordered particle lacks a velocity. Why %%@
should the unique realizer of that functional role have the causal properties and %%@
relations Tooley intuitively wants?\footnote{Bigelow and Pargetter's argument for %%@
the same conclusion, that an object can have orthodox velocity without Tooleyan %%@
velocity, is that movie-images and spots of  light have orthodox velocity but no %%@
Tooleyan velocity, for lack of an appropriate causal link (1989, p. 293-294). I %%@
reply: you are at liberty to introduce a notion of velocity stronger than the %%@
orthodox one by requiring some causal link, and your notion may mesh better with %%@
everyday use of `velocity'. But that hardly counts as a criticism of the orthodox %%@
notion.}

(4): {\em Tooley's Section 4.6}:--- \\
In Section 4.6, the imagined discontinuous motion is much more extreme. Tooley %%@
asks us to 
\begin{quote}
consider the following case. The world contains a rather unusual force field that %%@
causes objects to ``flash'' in and out of existence. Specifically, any object [it %%@
had better be a point-particle!---JNB] that enters this field exists only at %%@
points whose distance from the center of the field, measured in terms of a certain %%@
privileged unit of length, is given by an irrational number. Thus, if an object is %%@
moving along, and enters the field, it blinks in and out of existence an infinite %%@
number of times in any interval, however short. ... Given that the object is %%@
progressing through the field, it is natural to describe it as being in motion. %%@
Moreover, given that Achilles would, we can suppose, pass through a given field %%@
more quickly than the tortoise, it seems natural to say that objects have %%@
different velocities as they move through the field ... [But] the standard account %%@
of velocity [will not] assign a velocity to objects that are flashing along %%@
through [this] peculiar force field. In contrast, if velocity is a theoretical %%@
property of an object at a time, there would seem to be no reason why objects %%@
could not possess a velocity as they move along, blinking in and out of existence. %%@
(p. 247-248) 
\end{quote}
In reply, I think many philosophers of physics will find this thought-experiment %%@
so physically unrealistic that they will be happy to discount any ``intuitions'' %%@
about whether there is motion and velocity in it. Fair enough, say I. But there is %%@
also a more specific response, which builds on my earlier comments.\\
\indent We noted at the end of Section \ref{ssec;vely?} that the orthodox  account %%@
of velocity could be readily generalized to attribute velocity at a time that was %%@
a limit point (or even a one-sided limit point) of a domain of definition of a %%@
position function $q$. In Tooley's thought-experiment, this is of course exactly %%@
what occurs---all the time! Thus Tooley is presumably imagining the simplest sort %%@
of case where the discontinuous worldline of the particle is a dense subset of a %%@
smooth curve in spacetime. To put the case heuristically in terms of %%@
counterfactuals: the particle would have had the smooth curve as its worldline, %%@
and so an orthodox velocity, were it not being ``flashed out of existence'' when %%@
at rational distances from the center of the field. For this sort of case, the %%@
orthodox account can be readily generalized: the discontinuous worldline %%@
determines a unique smooth curve, so that at each time when the particle exists it %%@
can be attributed unambiguously the velocity associated with that point on the %%@
curve. (The same comment could of course be made using the discontinuous spatial %%@
trajectory  rather than the worldline.)\\
\indent Again, there are good mathematical questions hereabouts: viz. about the %%@
conditions under which ``bad'' curves, e.g. discontinuous ones, have ``good'' e.g. %%@
differentiable extensions. But Tooley does not pursue these questions; and nor %%@
will I. He just says, as I quoted above, that his kind of account of velocity as a %%@
theoretical property would surely attribute a velocity to the ``flashing'' %%@
particle. So be it, say I. But I deny that this merit is thanks to features %%@
specific to Tooley's account (such as  velocity being a cause or effect, or being %%@
functionally defined). The reason his account could, or would, attribute velocity %%@
is the simple mathematical one: that even the orthodox account can easily be %%@
generalized to  do this; and since Tooley's $T_1 \& T_2$ is a mathematically mild %%@
generalization of the orthodox account, it also can be thus generalized. Nothing %%@
specific to Tooley's account seems relevant.\footnote{In Section 6.1, Tooley %%@
admits that for the particle's velocity function, which is defined on the dense %%@
subset of the worldline, to be integrable, as $T_1$ demands, his account needs to %%@
interpret $T_1$ as using the Lebesque rather than Riemann theory of integration. %%@
Fair comment: but the orthodox account can equally adopt Lebesque integration.}

\subsubsection{Tooley's further discussion}\label{sssec;Tooleymore}
I turn to Tooley's further discussion in his Sections 5, 6. I shall consider two %%@
of his topics. (Footnote 24 replied {\em en passant} to a third.)

(1): {\em Tooley's Section 5.1}:--- \\
The first is his account's explanation of why the orthodox account works as well %%@
as it does (his 5.1). Tooley thinks it likely that in the actual world the two %%@
accounts will always coincide, in the sense that the orthodox and Tooleyan %%@
velocities will:\\
\indent (a) be defined in all the same cases and \\
\indent (b) be equal.\\
He argues for (a) by saying:\\
\indent (i): $T_2$ `ensures that an object's velocity at a time is causally %%@
relevant to its velocity at later times, and this means that the sort of situation %%@
where an object has an [orthodox] velocity, but fails to have a [Tooleyan] %%@
velocity---namely, cases of accidentally orderly movement---cannot arise in our %%@
world' (p. 249).\\
\indent (ii): On both accounts, velocity changes require forces. This
\begin{quote} ... together with plausible hypotheses about how forces depend upon %%@
other factors, such as the distance between the objects in question, entails that %%@
velocity, in our world, cannot change in a discontinuous fashion. Accordingly, the %%@
sorts of cases where an object can have a [Tooleyan] velocity, but fail to have an %%@
[orthodox] velocity [i.e. cases like the particle which is at rest and then moves %%@
with velocity $v(t) = 1$ for all $t > 0$, in his Section 4.5.---JNB], cannot occur %%@
in our world (p. 249).
\end{quote} 
Finally Tooley argues for (b) by saying that since both orthodox velocities and %%@
his velocities satisfy $T_1$, they must be equal whenever both are defined.

In reply: I applaud Tooley's seeking an argument why the orthodox account works as %%@
well as it does: better an argument than just postulating a law of nature that the %%@
two accounts always coincide.\footnote{As Arntzenius (2000: p. 196) and Bigelow %%@
and Pargetter (1989, p. 294) do: admittedly, in much briefer discussions.} But I %%@
find Tooley's argument unpersuasive as regards (a) and (b): only a much more %%@
elaborated theory of causation and motion could sustain the inferences needed. \\
\indent Thus, as to (i): why believe that accidentally orderly movement is the %%@
only way to have an orthodox but not Tooleyan velocity?. Tooley's (ii) obviously %%@
does not purport to be more than a sketch. But whatever the `plausible hypotheses' %%@
might be, there are basic problems about the strategy of the argument. Why should %%@
discontinuous changes in velocity be the only way to have a Tooleyan velocity but %%@
not an orthodox one? After all, the latter requires non-differentiability of %%@
position, not discontinuity of its derivative. Besides, discontinuous changes in %%@
forces mean discontinuities in acceleration: which need not spell discontinuities %%@
in velocity.\\
\indent Finally, even if (a) {\em were} established, (b) would not follow just %%@
from the fact that both orthodox and Tooleyan velocities satisfy $T_1$. After all,  %%@
integration is an averaging operation and so ``loses information''. So even if %%@
$q(t)$ is differentiable, so that indeed $q = \int dq/dt \; dt$, there are still %%@
countless other functions $v \neq dq/dt$ such that $q = \int v \; dt$: for %%@
example, $v$ could be a ``scarring'' of a smooth $dq/dt$ by inserting some %%@
discontinuities.

(2): {\em Tooley's Section 6.2}:--- \\
The second topic is the problem I emphasised at the outset (Section %%@
\ref{sssec;commonview}.B) as confronting all advocates of Tooleyan velocities: the  %%@
conflict between Tooleyan velocities being intrinsic, and velocity being relative %%@
to a frame of reference.\\
\indent Tooley treats this briefly in his Section 6.2 (p. 251). He sees it as a %%@
matter of reconciling his proposal with special relativity. He admits that %%@
`philosophical criticisms of scientific  theories do not have an extraordinarily %%@
impressive track record', but conjectures that a heterodox interpretation of %%@
special relativity which adds a relation of absolute simultaneity may be tenable, %%@
for example because it better allows a tensed account of the nature of time. (He %%@
develops this conjecture in his (1997); cf footnote 12.)

\indent In reply, I will not repeat Section \ref{sssec;commonview}.B's endorsement %%@
of the problem. But I would make two further comments, specifically about Tooley's %%@
answer to it. First, I would be much less willing than Tooley to let metaphysical %%@
views determine the interpretation of physical theories, and in particular to %%@
adopt a tensed account of time. Second I emphasise that relativity's denial of %%@
absolute simultaneity is irrelevant to the problem. As noted in (1) of Section %%@
\ref{sssec;commonview}.B, classical mechanics no less than relativity can be %%@
formulated with velocity being relative to a frame of reference. So the problem %%@
arises already on Tooley's chosen ``home-ground'' of classical mechanics.

So to sum up this and the previous Subsection's critique of Tooleyan velocities: I %%@
have argued that Tooley and other authors (i) do not justify crucial premises of %%@
their arguments for these velocities, and (ii) under-estimate the resources of the %%@
orthodox account.

\section{``Shadow velocities'': Lewis and Robinson}\label{sssec;LewRob}
I turn to a proposal of Lewis and Robinson that is similar in some ways to Section %%@
\ref{sssec;Tooley}'s proposed intrinsic velocity. In short, they propose that a %%@
moving object has a vectorial property (i.e. a property represented by a vector) %%@
which is intrinsic to the object, and whose vector is equal to the velocity %%@
vector. But this property is not itself velocity: hence this Section's title. For %%@
velocity presupposes the persistence of the moving object;  and this property is %%@
to be intrinsic, not merely (as we have emphasised, especially in Section %%@
\ref{ssec;vely?}) ``almost intrinsic''. In fact, Robinson (1989) floats the %%@
proposal but does not endorse it; Lewis (1999) endorses it. (As we shall see, this %%@
difference between them turns on our central question,  familiar since Section %%@
\ref{ssec;pismprospect}: can vectorial properties be intrinsic to a point?)

\indent But unlike Tooley, Lewis and Robinson are concerned about persistence. %%@
Specifically, they take their proposal to provide the perdurantist, especially the %%@
advocate of Humean supervenience (Lewis 1986, pp. ix-x; 1994, pp. 225-226), with a %%@
reply to the rotating disc argument. I discuss persistence as a context for their %%@
proposal elsewhere (2004, Section 4.3). So in this paper, I will mostly leave %%@
aside this context:  my pro-perdurantist claim (FPe) in Sections %%@
\ref{sssec;my2claims} and \ref{sssec;forperdm}  will suffice. (I also leave aside %%@
how even a simple quantity such as mass defined on the points and regions of a %%@
continuous body causes trouble for {\em pointillisme}, in particular Humean %%@
supervenience: cf. Butterfield (2006) or Hawthorne (2006: Section 2).)

I will first present the proposal, in Section \ref{sssec:LRproposal}.  Then in %%@
Section \ref{sssec;LRassessed} I will criticize it. Then the last two Subsections %%@
try to offer a peace-pipe to Lewis and Robinson. In Section \ref{secIntro}, I use %%@
Hilbert's $\varepsilon$ operator to define a quantity which is like their proposed %%@
quantity, in being analogous to velocity yet not presupposing persistence. I will %%@
call it `welocity'. Finally in Section \ref{332B:whatpricehs}, I briefly ask %%@
whether welocity satisfies Lewis' and Robinson's goals: but again my conclusion is %%@
negative---it would not.

\subsection{The proposal}\label{sssec:LRproposal}
Recall the rotating disc argument. The perdurantist is challenged to say what %%@
distinguishes two rigid congruent utterly homogeneous discs, one rotating and one %%@
stationary. It seems that the perdurantist, with her meagre resources, in %%@
particular ``qualitative'' facts intrinsic to temporal instants, cannot do so: she %%@
cannot ``thread the worldlines together'' correctly.  

Robinson (1989; p. 405 para 2, p. 406 para 2 to p. 408 para 1) floats the %%@
following reply. It combines the idea of a ``cousin'' of velocity that is %%@
intrinsic and does not presuppose persistence, with the idea (endorsed by many %%@
philosophers) that persistence is determined (``subvened'') by relations of %%@
qualitative similarity and causal dependence between events. To be precise, the %%@
reply  has five components; as follows:\\
\indent (i) a vectorial property at a point can be an intrinsic property of that %%@
point;\\
\indent (ii) the propagation of continuous matter through spacetime  involves such %%@
a property at every spacetime point; and \\
\indent (iii) these properties distinguish the rotating and non-rotating discs, %%@
since the vector that represents the property at a point is timelike, and  points %%@
in the same direction as the instantaneous four-dimensional velocity vector at %%@
that point; \\
\indent (iv) the distribution of these properties, from point to point, determines %%@
(subvenes) the relations of qualitative similarity between points, and especially %%@
the relations of causal dependence between events at those points; and\\
\indent (v) the distribution of these properties, by determining the lines of %%@
causal dependence, determines the lines of persistence.

But Robinson himself has second thoughts about this proposal. His doubts concern %%@
the first component, (i): i.e. the Yes answer to Section \ref{ssec;pismprospect}'s %%@
question. He thinks the directionality of a vector forbids it from representing an %%@
intrinsic property; and he backs this up with an argument about a point and a %%@
duplicate of it, which he credits to Lewis in discussion.\footnote{Similar doubts %%@
are expressed by other contemporary metaphysicians, some sharing a Lewisian %%@
approach to the intrinsic-extrinsic distinction. Butterfield (2006: Section 4.1) %%@
gives references.}

On the other hand, Lewis believed (at least by about 1993) that vectorial %%@
properties could be intrinsic to points (1994, p. 226). And in a final short paper %%@
on the rotating disc argument (replying to Zimmerman's critique of perdurantism, %%@
1998), Lewis endorsed Robinson's proposal; (Lewis 1999, p. 211).\footnote{My %%@
(2004, Section 4.3.1) gives more details about how Lewis came around  (ca. 1998) %%@
to this proposal, after espousing for a while (ca. 1986-1994) a more %%@
``stone-walling'' reply.\\
\indent Note also that the proposal is  clearly similar in spirit to Tooley's  %%@
heterodoxy about velocity, as in Section \ref{sssec;Tooley}. Robinson does not %%@
refer to Tooley et al.: the work was of course  contemporaneous. But Zimmerman %%@
(1998, p. 281, p. 284) and Sider (2001, p. 228) both see the similarity. Zimmerman %%@
first discusses reading Robinson's proposal as the same as Tooley's (p. 281), and %%@
then discusses reading it as just similar (p. 284, note 65). Sider reads the %%@
proposals as  similar. More specifically, Sider and Zimmerman's second reading %%@
both see Robinson's proposal as going with an orthodox, or ``Russellian at-at'', %%@
account of motion. So also (implicitly) does Lewis' discussion.}   

So the idea of the proposal is that the difference in the properties  postulated %%@
by (ii) and (iii) amounts to a difference in the `local arrangement of qualities' %%@
as demanded by Humean supervenience. Thus Lewis (1999, p. 211) begins by %%@
approvingly quoting Robinson, suggesting we should
\begin{quote} 
\ldots see the collection of qualities characteristic of the occupation of space %%@
by matter as in some sense jointly self-propagating; the fact of matter occupying %%@
space is itself causally responsible ... for the matter going on occupying space %%@
in the near neighbourhood immediately thereafter. ... [The posited vectors] figure %%@
causally in determining the direction of propagation of [themselves as well as] %%@
other material properties. (Robinson 1989, p. 406-407.) 
\end{quote}
Lewis then goes on to formulate the proposal more formally, as a putative law that %%@
partially specifies a vector field $V$. The specification is partial, both in (i) %%@
being admitted to be a ``first approximation'', and (ii) specifying only the %%@
direction but not the length of the vector at each point. But (ii) hardly matters: %%@
it will be obvious that Robinson and Lewis could frame their proposal entirely in %%@
terms of postulating a timelike direction field (i.e. a specification at each %%@
point of continuous matter of a timelike direction), rather than a vector field. %%@
But I shall follow them and talk of a vector field.\\
\indent In giving this formulation, Lewis' aim is partly to avoid various %%@
objections or limitations. In particular, the formulation should not invoke either %%@
persistence or causation, since  these are meant to supervene on the local %%@
arrangement of qualities, taken of course as including facts about the vector %%@
field $V$. Thus the formulation is to avoid circularity objections that had been %%@
urged by Zimmerman (1998) against some related proposals.\\
\indent So in particular: the vector field $V$ cannot simply be the instantaneous %%@
(four-dimensional) velocity (orthodox, not Tooleyan!) of the matter at the point %%@
in question.  For $V$ is to contribute to an analysis of (or at least to a %%@
supervenience basis for) persistence and thereby of velocity.\\
\indent Similarly, since Lewis agrees that causation is crucial to persistence %%@
(`the most important sort of glue that unites the successive stages of a %%@
persisting thing is causal glue': 1999, p. 210), causation cannot be invoked in %%@
the course of specifying the vector field $V$. 

Lewis proposes that (for a world with continuous space and time), the %%@
specification of $V$ `might go something like this':
\begin{quote}
Let $p$ be any spacetime point, and let $t$ be any smooth timelike trajectory %%@
through spacetime with $p$ as its final limit point. Let each point of $t$ before %%@
$p$ be occupied by matter with its vector [i.e. vector of the vector field $V$] %%@
pointing in the direction of $t$ at that point. [So in the jargon of modern %%@
geometry, $t$ is an integral curve of $V$.] Then, {\em ceteris paribus}, there %%@
will be matter also at $p$. (1999, p. 211.)
\end{quote}
Here, the `{\em ceteris paribus}' clause is to allow for the fact that the %%@
point-sized bit of matter might cease to exist before $p$, because of `destructive %%@
forces or self-destructive tendencies' (ibid.).\\
\indent Lewis also stresses that this proposal is to be read as a law of %%@
succession, not of causation. This means, I take it, that the `Then, {\em ceteris %%@
paribus}' is to be read as a material conditional.

\subsection{Criticism: the vector field remains %%@
unspecified}\label{sssec;LRassessed}
I claim that Lewis' proposal fails. It is too weak: it does not go far enough to %%@
specify $V$. For it only says, of any timelike open curve that is an integral %%@
curve of $V$, that the future end-point $p$ of this curve will, {\em ceteris %%@
paribus}, have matter at it.\\
\indent But every suitably smooth vector field $U$ defined on a open region $R$ of %%@
spacetime has integral curves throughout $R$; (which are timelike, by definition, %%@
if $U$ is). (To be precise: `suitably smooth' is none too demanding: all we need %%@
is that $U$ be $C^1$, i.e. the partial derivatives of its components exist and are  %%@
continuous.) So suppose Lewis stipulates, that the field $V$ is to be timelike and %%@
$C^1$ on an open set $R$ which is its domain of definition (say, the %%@
spatiotemporal region occupied by continuous matter): which (``giving rope'') we %%@
can assume to be a legitimate, in particular non-circular, stipulation. Then his %%@
proposal says that, {\em ceteris paribus}, every point $p \in R$ has matter at %%@
it.\\
\indent But that claim hardly helps to distinguish $V$ from the countless other %%@
(timelike smooth) vector fields $U$. For however exactly one interprets `{\em %%@
ceteris paribus}', the claim is surely true of $p$ regardless of the integral  %%@
curve one considers it as lying on. So the claim about $p$ does not constrain the %%@
vector field.  Indeed, if Lewis sets out to specify $V$ on the spatiotemporal %%@
region occupied by continuous matter, the claim is thereby assumed to be true for %%@
all  $p$ in the region, regardless of vector fields. So again, we have said %%@
nothing to distinguish $V$ from the countless other vector fields $U$.\\
\indent Agreed, Lewis puts forward his proposal as a ``first approximation'' to %%@
specifying $V$. But so far as I can see, his discussion doesn't contain any %%@
ingredients which would, for continuous matter, help distinguish $V$ from other %%@
vector fields.\footnote{Nor can I guess how I might have misinterpreted Lewis' %%@
proposal. The situation is puzzling: and not just because Lewis thought so %%@
clearly, and my objection is obvious. Also, the objection is analogous to what %%@
Lewis himself says (p. 210) against the naive idea that $V$ should point in the %%@
direction of perfect qualitative similarity: viz. that `in non-particulate %%@
homogeneous matter, ... lines of qualitative similarity run every which way'.

An anonymous referee suggests that Lewis' attempted specification of $V$ might %%@
succeed if the property is required to be natural, in Lewis' sense; or at least, %%@
might succeed if the environment around the region $R$ is also sufficiently %%@
heterogeneous. I confess I do not see how naturalness and-or a heterogeneous %%@
environment will help Lewis; but I discuss the latter in the sequel.}

I should note here that Zimmerman (1999) makes a somewhat similar objection to %%@
Lewis' proposal. But his exact intent is not clear to me.\\
\indent He maintains  that in some seemingly possible cases of continuous matter, %%@
Lewis' proposal does not specify a unique vector field $V$---indeed hardly %%@
constrains $V$ at all.  He says (p. 214 para 1 and 2) that in possible worlds with %%@
a  physics of the sort Descartes might have envisaged, i.e. where there is {\em %%@
nowhere} any vacuum, and only {\em one} kind of (continuous homogeneous) stuff %%@
fills all of space: `{\em every} vector field will satisfy [Lewis'] law.'\\
\indent Thus Zimmerman assumes that:\\
\indent \indent (i): the worlds with which he is concerned are wholly filled with %%@
the one kind of stuff; and\\
\indent \indent (ii): these worlds are thus filled as a matter of law, not %%@
happenstance (in the jargon: as a matter of physical or nomic necessity).\\
\indent He also says (p. 214-5) that he needs to assume (i) and (ii) in order to %%@
criticize Lewis' proposal, together with obvious modifications of it which allow %%@
for different types (``colours'') of continuous matter. That is: Zimmerman thinks %%@
Lewis' proposal works, or could be modified to work, for worlds in which:\\
\indent (i') continuous matter  does not fill all of space and-or comes in various %%@
types; or\\
\indent (ii') continuous matter of just one type fills all of space, but only as a %%@
matter of happenstance. 

\indent In view of my own objection, I do not understand why Zimmerman feels he %%@
needs to assume (i) and (ii) in order to object to Lewis. He does not explicitly %%@
say why he does so. {\em Maybe} it is to block some Lewisian rejoinder, that would %%@
better specify $V$, by adding constraints of either or both of two kinds:\\
\indent (i''): constraints about the spatiotemporal relations of the continuous  %%@
matter  in a bounded volume (say, one of our discs) to other matter outside the %%@
volume;\\
\indent (ii''): constraints about the nomic or modal properties of matter.\\
But it remains unclear how the details of (i'') and (ii'') might go. 

To sum up: For all I can see, my objection, that $V$ is not distinguished from %%@
countless other vector fields, applies to Lewis' proposal (and thereby: the spirit %%@
of Zimmerman's objection also applies) for the case that Lewis intended it---i.e. %%@
the discs of the original rotating discs argument.

\subsection{Avoiding the presupposition of persistence, using Hilbert's %%@
$\varepsilon$ symbol}\label{secIntro}
Since Section \ref{sssec;Tooley}, my discussion has been critical. Now I try to be %%@
more constructive! I propose to make precise the idea that velocity, understood in %%@
the orthodox way, is hardly extrinsic. Elsewhere (2004, Section 4.2.2; 2004a, %%@
Section 4.5) I make this precise in two related ways. Here I develop a third %%@
way.\\
\indent Namely, I will define a quantity which will be like Robinson and Lewis' %%@
proposal from Section \ref{sssec:LRproposal}, in that it is (``usually'') equal to %%@
instantaneous velocity, and yet does {\em not} presuppose persistence. But unlike %%@
their proposal, it has no {\em pointilliste} motivations: in fact, it is adapted %%@
from the orthodox definition of velocity.  After presenting it, I will end by %%@
comparing it with their proposal (Section \ref{332B:whatpricehs}). 

\indent I will call my new-fangled quantity {\em welocity}, the `w' being a %%@
mnemonic for `(logically) weak' and-or `without (presuppositions)'. So my goal is %%@
that welocity is to reflect, in the way its values are defined, this lack of %%@
presupposition. That is: the values are to be defined in such a way that it is %%@
impossible to infer from the value of the welocity of the object $o$ at time $t$ %%@
that $o$ in fact exists in a neighbourhood of $t$, and has a differentiable %%@
worldline at $t$: an inference which, as we have just seen, {\em can} be made from %%@
the value of velocity as orthodoxly understood.\\
\indent There are three preliminary points to make about this goal: the first %%@
philosophical,  the second mathematical and the third physical. But only the first %%@
represents a limitation of scope for what follows.\\
\indent  First, presupposition is often taken to be a subtler notion than just %%@
`necessary condition'. So I admit that for a property ascription to lack a %%@
presupposition of persistence, it is perhaps {\em not} enough that the ascription %%@
fails to imply that the instance persists. But I will set this aside: I will aim %%@
only for the modest goal of avoiding the implication (if not, perhaps, the %%@
presupposition {\em stricto sensu}) of persistence.\\
\indent Second, I just said that $o$'s having an orthodox velocity at $t$ implies %%@
$o$'s existing in an open neighbourhood of $t$, and its position in space ${\bf %%@
q}(t)$ being differentiable at $t$. Indeed, that is orthodoxy. But  as I remarked %%@
at the end of Section \ref{verdict}, one can generalize so as to imply only that %%@
$o$ exists at a set of times for which $t$ is a limit point (and that its average %%@
velocities go to a common limit at $t$). But to avoid cumbersome phrasing, I will %%@
from now on not repeat this generalization.\\
\indent Third, as I stressed (against Tooley and others) in Sections %%@
\ref{sssec;commonview} and \ref{sssec;Tooleymore}: orthodox velocity is relative %%@
to a frame of reference. And one cannot expect that avoiding an implication of %%@
$o$'s persistence will also make for avoiding the implication of, and relativity %%@
to, a frame. Agreed: and indeed, ascriptions of welocity will be just as obviously %%@
relative to a frame (and so in that way extrinsic) as are ascriptions of velocity. %%@
But in order to engage better with authors such as Tooley, Robinson and Lewis, I %%@
will not emphasise this aspect.

\indent Developing this idea---values of a quantity like velocity, but which do %%@
not imply $o$'s persistence nor its worldline's differentiability---takes us to a %%@
familiar philosophical territory: viz., rival proposals for the semantics of empty %%@
referring terms. In our case, the empty terms will be expressions for $o$'s %%@
instantaneous velocity at $t$; and, as just discussed, they can be empty because  %%@
either:\\
\indent \indent (NotEx): $o$ does not exist for an open interval around $t$, or\\
\indent \indent (NotDiff): $o$ does exist for an open interval around $t$, but its %%@
position ${\bf q}$ is not differentiable at $t$; (roughly: there is a ``sharp %%@
corner'' in the worldline).\\
\indent (And similarly for acceleration and higher derivatives; but I shall %%@
discuss only velocity---tempting though words like `wacceleration' are!)

\indent In fact, it will be clearest to lead up to my proposal for welocity  by %%@
first considering a simpler one, which is modelled on Frege's proposal that (to %%@
prevent truth-value gaps) empty terms should be assigned some %%@
``dustbin-referent'', such as the empty set $\emptyset$. Thus if one sets out to %%@
define a quantity that is like velocity but somehow avoids its presupposition of %%@
persistence, one naturally first thinks of a quantity, call it $\bf u$, defined to %%@
be\\
\indent \indent (a): equal to the (instantaneous)  velocity $\bf v$, for  those %%@
times $t$ at which $o$ {\em has} a velocity; and\\
\indent \indent  (b):  equal to some dustbin-referent, say the empty set %%@
$\emptyset$, at other times $t$; i.e. times such that either (NotEx): $o$ does not %%@
exist for an open interval around $t$; or (NotDiff): $o$ does exist for an open %%@
interval around $t$, but its position ${\bf x}$ is not differentiable at $t$.\\
\indent Of course,  variations on (b) are possible. One could select different %%@
dustbin-referents for the two cases, (NotEx) and (NotDiff), (say, $\emptyset$ and %%@
$\{\emptyset \}$) so that $\bf u$'s value registered the different ways in which %%@
an instantaneous  velocity could fail to exist. And instead of using a %%@
dustbin-referent, one could say that the empty term just has no ``semantic %%@
value'', or ``is undefined'': (a contrast with dustbin-referents which would %%@
presumably show up in truth-value gaps, and logical behaviour in general).

Agreed, this definition is natural. But it does not do the intended job. For this %%@
quantity $\bf u$, whether defined using (b) or using the variations mentioned, %%@
does not avoid, in the way intended, the presupposition of persistence. For $\bf %%@
u$'s value (or lack of it, if we take the no-semantic-value option) registers %%@
whether or not the presupposed persistence holds true. That is: we {\em can} infer %%@
from the value of $\bf u$ (or its lack of value) whether (a) $o$ has a velocity in %%@
the ordinary  sense, or (b) the presupposition has failed in that (NotEx) or %%@
(NotDiff) is true. In short: $\bf u$'s individual values tell us too much.

But there {\em is} an appropriate way of assigning semantic values to empty terms, %%@
i.e. a way of defining a quantity, {\em welocity}, that is like velocity but whose %%@
values do not give the game away about whether the presupposition has failed, i.e. %%@
about whether (NotEx) or (NotDiff) is true. In order not to give the game away, %%@
welocity must  obviously take ordinary values, i.e. triples of real numbers %%@
(relative to some frame of reference), even when the presupposition has failed. %%@
But how to assign them?\\
\indent The short answer is: arbitrarily. The long answer is: we can adapt schemes %%@
devised by logicians in which a definite description, whose predicate has more %%@
than one instance, is assigned as a referent {\em any one} of the objects in the %%@
predicate's extension. (The first such scheme was devised by Hilbert and Bernays; %%@
but we will only need the general idea.) Such a scheme applies to our case, %%@
because we can write the definition of welocity in such a way that when the %%@
presuppositions fail (i.e. (NotEx) or (NotDiff) is true), the   predicate (of %%@
triples of real numbers) in the definition  is vacuously satisfied by {\em all} %%@
such triples; so that forming a definite description, and applying semantic rules %%@
like Hilbert-Bernays', welocity  is assigned an {\em arbitrary} triple of real %%@
numbers as value.\\
\indent Thus we get the desired result: if you are told that the value of welocity  %%@
for $o$ at $t$ is some vector in $\mathR^3$, say (1,10,3) relative to some axes %%@
and choice of a time-unit, you cannot tell whether:\\
\indent (a): (NotEx) and (NotDiff) are both false (i.e. the presuppositions of %%@
velocity hold), and $o$ has velocity (1,10,3); or\\
\indent (b): Either (NotEx) or (NotDiff) is true, the predicate is vacuously %%@
satisfied by all triples, and (1,10,3) just happens to be the triple assigned by %%@
semantic rules taken from Hilbert-Bernays' (or some similar) scheme.

The details are as follows. (1): Hilbert and Bernays introduced the notation %%@
$(\varepsilon x)(Fx)$ for the definite description `the $F$', with the rule that %%@
if $F$ had more than one instance, then $(\varepsilon x)(Fx)$ was assigned as %%@
referent any such instance, i.e. any element of $F$'s extension. (We need not %%@
consider their rule in more detail than this; nor their rule for what to say when %%@
$F$ has no instances; nor their rules' consequences for the semantics and syntax %%@
of singular terms. For details, cf. Leisenring (1969).)

\indent (2): Next, we observe that the {\em velocity} of an object $o$ at time $t$ %%@
relative to a given frame can be defined with a definite description containing a %%@
material conditional whose antecedents are the presuppositions of persistence and %%@
differentiability. That is: velocity can be defined along the following lines:---

\noindent The velocity of $o$ at time $t$ (relative to a given frame) is the %%@
triple of real numbers ${\bf v}$ such that:
\begin{quote}
for some (and so any smaller) open interval $I$ around $t$:\\
\{[$o$ exists throughout $I$]  and [$o$'s position $\bf x(t)$ is differentiable in %%@
$I$]\} $\;\;\; \supset \;\;\;\;$ [$\bf v$ is the common limit of average %%@
velocities for times $t' \in I$, compared with $t$, as $t' \rightarrow t$ from %%@
above or below].
\end{quote}
This {\em definiens} uses a material conditional. So it will be vacuously true for %%@
all triples $\bf v$, if the antecedent is false for all open intervals $I$ around %%@
$t$, i.e. if (NotEx) or (NotDiff) is true: in other words, if velocity's %%@
presuppositions of continued existence and differentiability fail.

\indent (3): Now we put points (1) and (2) together. Let us abbreviate the %%@
displayed {\em definiens}, i.e. the open sentence with $\bf v$ as its only free %%@
variable, as $F({\bf v})$. Then I propose to define the {\em welocity}  of $o$ at %%@
$t$ by the singular term $(\varepsilon {\bf v})(F{\bf v})$: which is, by %%@
Hilbert-Bernays' semantic rule:\\
\indent \indent (a): equal to the (instantaneous)  velocity of $o$, for  those %%@
times $t$ at which $o$ {\em has} a velocity; and\\
\indent \indent  (b):  equal to some arbitrary triple of real numbers, at other %%@
times $t$; i.e. at times such that either (NotEx): $o$ does not exist for an open %%@
interval around $t$; or (NotDiff): $o$ does exist for an open interval around $t$, %%@
but its position ${\bf x}$ is not differentiable at $t$.

Welocity, so defined, has the desired features: its values do not give the game %%@
away about whether (NotEx) or (NotDiff) is true.

That is all I need to say about welocity, for this paper's purposes; and in %%@
particular,  for Section \ref{332B:whatpricehs}'s comparison with Robinson's and %%@
Lewis' proposal.\\
\indent But I end this Subsection by noting that there are of course various %%@
technical questions hereabouts, even apart from the logical questions about the %%@
$\varepsilon$ symbol. For example, a natural question arises from letting $o$ be a %%@
point-sized bit of matter in a continuum, and letting the presuppositions of %%@
velocity fail for various such bits of matter: some such bits may fail to exist, %%@
and some may have a non-differentiable worldline. One then asks: how widely across %%@
space, and in how arbitrary a spatial distribution, can these bits fail to exist, %%@
or have a non-differentiable worldline---i.e. how widely and arbitrarily can the %%@
presuppositions of velocity fail---while yet the welocity field might not ``give %%@
the game away'', in that the arbitrary values {\em can} be assigned at all the %%@
points  where the presuppositions  fail, so as to give a smooth (e.g. continuous %%@
or even differentiable) welocity field? This is in effect a question about the %%@
scope and limits of regularization of singularities in real vector fields: a good %%@
question---but not one for this paper!

\subsection{Comparison with Robinson and Lewis}\label{332B:whatpricehs}
I shall compare welocity with Robinson's and Lewis' proposal in two stages; the %%@
first more specific than the second.

\indent (1): {\em Is welocity intrinsic?}:---\\
In the light of Section \ref{ssec;pismprospect}'s central question, this is the %%@
obvious question to ask about welocity! But in asking this, I will set aside the %%@
fact that welocity, like velocity, is relative to a frame. This sets aside the %%@
existence of other objects representing the frame, and focuses attention on %%@
extrinsicality arising from implications about the other temporal parts of $o$ %%@
itself. This tactic will make for a less cumbersome comparison with Robinson and %%@
Lewis, and authors like Tooley; who, as we have seen,  tend to ignore the frame; %%@
(cf. the third preliminary point at the start of Section \ref{secIntro}).

The question whether welocity is intrinsic returns us to Section \ref{sssec;ied}'s %%@
idea of positive extrinsicality, i.e. the idea of implying accompaniment. Recall %%@
that according to Lewis (1983) and almost all succeeding authors,  this is a %%@
species of extrinsicality, since a property like being unaccompanied (more %%@
vividly: being lonely) is itself extrinsic; and that this species, positive %%@
extrinsicality, is agreed to be a good deal clearer than the genus, %%@
extrinsicality. Similarly for the negations: the negation of positive %%@
extrinsicality, i.e. not implying accompaniment (i.e. compatibility with being %%@
lonely), is weaker than---and a good deal clearer than---intrinsicality. So we %%@
might call it `weakened intrinsicality'. And as announced in Section %%@
\ref{sssec;ied}, my campaign against {\em pointillisme} can mostly take {\em %%@
pointillisme} to advocate weakened intrinsic properties.

Once we set aside frame-relativity and any extrinsicality ensuing from that, it is %%@
clear that both an ascription to $o$ at $t$ of some or other value of welocity, %%@
and an ascription of a specific value of welocity, are {\em not} positive %%@
extrinsic. For thanks to welocity allowing for (NotEx),  each ascription is %%@
compatible with $o$'s temporal part at $t$ being lonely: i.e. compatible with %%@
$o$'s not existing at other times. So the two ascriptions are weakened intrinsic. 

Besides, having some welocity or other (again setting aside frames and other %%@
objects) is a necessary property. For  $o$ has a welocity at $t$ iff: either\\
\indent (1) $o$ at $t$ is lonely, i.e.  (NotEx); or\\
\indent (2) $o$'s worldline is not differentiable at $t$, i.e.  (NotDiff);  or\\
\indent (3) $o$'s worldline is differentiable (and so $o$ has a velocity).\\
This disjunction is obviously equivalent to $o$'s merely existing at $t$; so that %%@
having some welocity or other is a necessary property.\\
\indent Now on Lewis' preferred analysis (1983a) and several alternatives (e.g. %%@
his ``fallback'' analysis in Langton and Lewis (1998))  an intrinsic property is %%@
one that does not differ between duplicate objects---where duplication is defined %%@
as sharing a certain elite minority of properties. Clearly, on any such analysis, %%@
any necessary property is intrinsic: for it will not differ between duplicates. %%@
(More generally: on any such analysis,  intrinsicality is not hyperintensional. %%@
That is, necessarily co-extensive properties are alike in being intrinsic, or %%@
not.)

 But on the other hand, having a {\em specific} welocity, say 5 ms$^{-1}$ North, %%@
is of course  not necessary. If this property applies to $o$ at $t$, while yet (1) %%@
and (2) are both false, then $o$ has a {\em velocity} 5 ms$^{-1}$ North. \\
\indent Indeed, I claim this property is extrinsic (though as just established, %%@
weakened intrinsic---setting aside frames). The argument is a general one; as %%@
follows.\\
\indent Consider a property $P$ defined along the lines: $o$ is $P$ iff: either %%@
(1') $o$ is lonely, or (2') if $o$ is accompanied, then $o$ and some accompanying %%@
objects satisfy some condition which is {\em not} necessary and which  involves %%@
them all ``non-redundantly''. Can we conclude that $P$ is intrinsic? Or extrinsic? %%@
Or can we make no conclusion?\\
\indent The intuitive verdict is surely that $P$ is extrinsic. Intuitions apart, %%@
$P$ is certainly extrinsic according to several analyses, e.g. by Lewis (1983a), %%@
Langton and Lewis (1998), Vallentyne (1997) and Lewis (2001). For example, for the %%@
first two analyses the reason is essentially that $P$ is a disjunction whose %%@
disjuncts are a positive extrinsic, viz. (2') and what Lewis (1983, p. 114) dubbed %%@
a negative extrinsic, i.e. a property implying loneliness, viz. (1').\\
\indent This common verdict suggests that not only welocity, but any quantity that %%@
is analogously defined with a  disjunct like (1'), will be extrinsic.

To sum up:--- We have established that having a specific value of welocity is a %%@
weakened intrinsic property; and that having some welocity or other is a necessary %%@
property, and so according to several analyses, intrinsic. On the other hand, %%@
having a specific value of welocity seems intuitively to be extrinsic; and %%@
certainly is extrinsic, according to several analyses.

\indent (2): {\em Rounding off}:---\\
To conclude: already in Sections \ref{widercpgn} and \ref{sec;pismintr}, I %%@
announced my denial of {\em pointillisme}, and so antipathy to Humean %%@
supervenience; and more specfically, my view that velocity was ``almost'' %%@
intrinsic. In Section \ref{secIntro}, we saw my peace-pipe for the {\em %%@
pointilliste}: at least, for the {\em pointilliste} who advocates weakened %%@
intrinsic properties. Namely, I defined a quantity, {\em welocity}, which is %%@
weakened intrinsic, and in that sense avoids velocity's implication of %%@
persistence. 

But I am afraid  Lewis would not smoke my peace-pipe!  For first, he would %%@
presumably be unimpressed by velocity's being almost intrinsic. For Humean %%@
supervenience is so central to his neoHumean metaphysical system that he sets %%@
great store by intrinsicality. So he would probably say that as regards failing to %%@
be intrinsic, a miss is as good (i.e. bad!) as a mile.

\indent Similarly, I expect that he would not welcome welocity. I agree that he %%@
might be ``envious''  of its being well-defined ({\em modulo} the freedom to %%@
assign referents associated with the $\varepsilon$ operator), since his own %%@
proposal, the intrinsic vector $V$, is yet to be successfully defined (Section %%@
\ref{sssec;LRassessed}). But Lewis is an advocate of intrinsicality, not just %%@
weakened intrinsicality; and as we have just seen, by Lewis' lights, welocity is %%@
extrinsic.

\vspace{1.0 truecm}

{\em Acknowledgements}:---  I am grateful to audiences at Florence, Kirchberg, %%@
Leeds, London, Oxford, and Princeton; and to F. Arntzenius, A. Elga, G. Belot, P. %%@
Forrest, J. Hawthorne, R. Le Poidevin, S. Leuenberger, J. Uffink, B. van Fraassen, %%@
D. Zimmerman and an anonymous referee, for helpful conversations and comments.

\newpage

\section{References}
Albert, D. (2000) {\em Time and Chance}, Harvard University Press.\\
F. Arntzenius (2000), `Are there really instantaneous velocities?', {\em The %%@
Monist} {\bf 83}, pp. 187-208.\\
F. Arntzenius (2003), `An arbitrarily short reply to Sheldon Smith on %%@
instantaneous velocities', {\em Studies in the History and Philosophy of Modern %%@
Physics} {\bf 34B}, pp. 281-282.\\
Arntzenius, F. (2004), `Time reversal operations, representations of the Lorentz %%@
group, and the direction of time', {\em Studies in the History and Philosophy of %%@
Modern Physics} {\bf 35B}, pp. 31-44.\\
Arthur, R. (2006) `Leibniz's syncategorematic infinitesimals, smooth infinitesimal %%@
analysis and Newton's proposition 6', forthcoming.\\
Batterman, R. (2003), `Falling cats, parallel parking and polarized light',  {\em %%@
Studies in the History and Philosophy of Modern Physics} {\bf 34B}, pp. 527-558. %%@
\\
Bell, J. (1998), {\em A Primer of Infinitesimal Analysis}, Cambridge University %%@
Press.\\
Belot, G. (1998), `Understanding electromagnetism', {\em British Journal for the %%@
Philosophy of Science} {\bf 49}, pp. 531-555.\\
Bigelow, J., Ellis, B. and Pargetter, R. (1988), `Forces', {\em Philosophy  of %%@
science} {\bf 55}, pp. 614-630.\\
Bigelow, J. and Pargetter, R.  (1989), `Vectors and Change', {\em British Journal %%@
for the Philosophy of Science} {\bf 40}, pp. 289-306.\\
Bigelow, J. and Pargetter, R.  (1990), {\em Science and Necessity}, Cambridge %%@
University Press.\\
Bricker, P. (1993), `The fabric of space: intrinsic vs. extrinsic distance %%@
relations', in ed.s P.French et al., {\em Midwest Studies in Philosophy} {\bf 18}, %%@
University of Minnesota Press, pp. 271-294.\\
Butterfield, J. (2004), `On the Persistence of Homogeneous Matter', available at: %%@
physics/0406021: and at http://philsci-archive.pitt.edu/archive/00002381/ \\
Butterfield, J. (2004a), `The Rotating Discs Argument Defeated', forthcoming in %%@
{\em British Journal for the Philosophy of Science}; available at:\\ %%@
http://philsci-archive.pitt.edu/archive/00002382/ \\
Butterfield, J. (2005)  'On the Persistence of Particles', in {\em Foundations of  %%@
Physics} {\bf 35}, pp. 233-269, available at: physics/0401112; and 
  http://philsci-archive.pitt.edu/archive/00001586/.\\
Butterfield, J. (2006), `Against {\em Pointillisme} about geometry', forthcoming %%@
in {\em Proceedings of the 28th Ludwig Wittgenstein Symposium}, ed. F. Stadler and %%@
M. St\"{o}ltzner; available at: http://philsci-archive.pitt.edu/archive. \\
Butterfield, J. (2006a), `Against {\em Pointillisme}: a call to arms', in %%@
preparation. \\
Colyvan, M and Ginzburg, L. (2003), `The Galilean turn in popoulation ecology', %%@
{\em Biology and Philosophy} {\bf 18}, pp. 401-414.\\
Dainton, B. (2001), {\em Time and Space}, Acumen Publishing.\\
Earman, J. (1987), `Locality, non-locality and action-at-a-distance: a skeptical %%@
review of some philosophical dogmas', in {\em Kelvin's Baltimore Lectures and %%@
Modern Theoretical Physics},  eds. R. Kargon and P. Achinstein, Cambridge Mass: %%@
MIT Press.\\
Earman, J. (1989), {\em World Enough and Spacetime}, MIT Press.\\
Earman, J. (2002), `What time reversal is and why it matters', {\em International %%@
Studies in the  Philosophy of Science} {\bf 16}, pp. 245-264.\\
Earman, J. and Roberts, J. (2006), `Contact with the Nomic: a challenge for %%@
deniers of Humean supervenience about laws of nature', {\em Philosophy and %%@
Phenomenological Research} forthcoming. \\
Hawthorne, J. (2006), `Quantity in Lewisian Metaphysics', forthcoming in his {\em %%@
Metaphysical Essays}, Oxford University Press.  \\
Hoefer, C. (2000), `Energy Conservation of in GTR', {\em Studies in the History %%@
and Philosophy of Modern Physics}, {\bf 31B} pp. 187-200. \\
Kingman, J. and Taylor, S. (1977), {\em Introducton to Measure and Probability}, %%@
Cambridge University Press.\\
Kragh, H. and Carazza, B. (1994), `From time atoms to spacetime quantization: the %%@
idea of discreet time 1925-1926', {\em Studies in the History and Philosophy of %%@
Modern Physics} {\bf 25}, pp. 437-462.\\ 
Langton, R. and Lewis, D. (1998), `Defining `intrinsic'', {\em Philosophy and %%@
Phenomenological Research} {\bf 58}, pp. 333-345; reprinted in Lewis (1999a), page %%@
reference to reprint.\\
Leibniz, G. (2001), {\em The Labyrinth of the Continuum: writings on the continuum %%@
problem 1672-1686}, ed. sel. and transl. R. Arthur, New Haven: Yale University %%@
Press.\\
Leisenring, A. (1969), {\em Mathematical Logic and Hilbert's $\varepsilon$ %%@
Symbol}, London: Macdonald Technical and Scientific.\\
Le Poidevin, R. (2006) `Motion, Cause and Moment', in preparation.\\
Lewis, D. (1970), `How to define theoretical terms', {\em Journal of Philosophy} %%@
{\bf 67}, pp. 427-446.\\ 
Lewis, D. (1983), `Extrinsic properties', {\em  Philosophical Studies} {\bf 44}, %%@
pp. 197-200; reprinted in Lewis (1999a); page references to reprint.\\
Lewis, D. (1983a), `New Work for a Theory of Universals', {\em Australasian %%@
Journal of Philosophy} {\bf 61}, pp. 343-77;  reprinted in Lewis (1999a).\\
Lewis, D. (1986), {\em Philosophical Papers, volume II}, New York: Oxford %%@
University Press.\\
Lewis, D. (1986a), {\em On the Plurality of Worlds}, Oxford: Blackwell.\\
Lewis, D. (1994), `Humean Supervenience Debugged', {\em Mind} {\bf 103}, p %%@
473-490; reprinted in Lewis (1999a), pp. 224-247; page reference to reprint.\\
Lewis, D. (1999), `Zimmerman and the Spinning sphere', {\em Australasian Journal %%@
of Philosophy} {\bf 77}, pp. 209-212.\\
Lewis, D. (1999a), {\em Papers in Metaphysics and Epistemology}, Cambridge:  %%@
University Press.\\
Lewis, D. (2001), `Redefining `intrinsic'', {\em Philosophy and Phenomenological %%@
Research} {\bf 63}, pp. 381-398.\\
Malament, D. (2004), `On the time reversal invariance of classical electromagnetic %%@
theory',  {\em Studies in the History and Philosophy of Modern Physics} {\bf 35B}, %%@
pp. 295-315.\\ 
Mancosu, P. (1996), {\em Philosophy of Mathematics and Mathematical Practice in %%@
the Seventeenth Century}, Oxford: University Press.\\
Robinson, A. (1996), {\em Non-standard Analysis}, Princeton University Press.\\
Robinson, D. (1989), `Matter, Motion and Humean supervenience', {\em Australasian %%@
Journal of Philosophy} {\bf 67},  pp. 394-409.\\
Russell, B. (1903), {\em The Principles of Mathematics}, London: Allen and %%@
Unwin.\\
Sider, T.  (2001), {\em Four-Dimensionalism}, Oxford University Press.\\
Sklar, L. (1974), {\em Space, Time and Spacetime}, University of California Press. %%@
\\
S. Smith (2003), `Are instantaneous velocities real and really instantaneous?', %%@
{\em Studies in the History and Philosophy of Modern Physics} {\bf 34B}, pp. %%@
261-280.\\
S. Smith (2003a), `Author's response', {\em Studies in the History and Philosophy %%@
of Modern Physics} {\bf 34B}, pp. 283.\\  
Tooley, M. (1987) {\em Causation: a Realist Approach}, Oxford: University Press.\\
M. Tooley (1988), `In Defence of the Existence of States of Motion', {\em %%@
Philosophical Topics} {\bf 16}, pp. 225-254.\\
Tooley, M. (1997) {\em Time, Tense and Causation}, Oxford: University Press. \\
Vallentyne, P. (1997), `Intrinsic properties defined', {\em Philosophical Studies} %%@
{\bf 88}, pp. 209-219.\\
Weatherson, B. (2002), `Intrinsic vs. extrinsic properties', {\em Stanford %%@
Encyclopedia of Philosophy}, http://plato.stanford.edu/intrinsic-extrinsic.\\
Wilson, M (1998), `Classical mechanics'; in {\em The Routledge Encyclopedia of %%@
Philosophy}.\\
Wilson, M. (2004), `Theory facades', {\em Proceedings of the Aristotelian Society} %%@
{\bf 78}, pp. 271-286.\\
D. Zimmerman (1998), `Temporal parts and supervenient causation: the %%@
incompatibility of two Humean doctrines', {\em Australasian Journal of Philosophy} %%@
{\bf 76}, pp. 265-288.\\
Zimmerman, D. (1999), `One really big liquid sphere: reply to Lewis', {\em %%@
Australasian Journal of Philosophy} {\bf 77}, pp. 213-215.

\end{document}